\documentclass[preprint,showpacs, showkeys,secnumarabic,amssymb,amsmath,amsfonts,aps, pra]{revtex4}

\usepackage{amsmath}
\usepackage{amsfonts,amssymb}
\usepackage{amsthm}
\begin{document}

\hoffset 1truecm
\hsize 14truecm
\baselineskip=22pt

\newcommand{\Zc}{\mbox{${\cal Z}_c$}}
\newcommand{\Nat}{{\cal I} \!\!\! {\cal N}}
\newcommand{\Ree}{{\Bbb R}}

\newcommand{\be}{\begin{eqnarray}}
\newcommand{\ee}{\end{eqnarray}}
%-----------------------------------------------------------------------
\newcommand{\ba}{\begin{array}}
\newcommand{\ea}{\end{array}}
%------------------------------------------------------------------------------
\newcommand{\bena}{\begin{eqnarray}}
\newcommand{\eena}{\end{eqnarray}}
%------------------------------------------------------------------------------
\newcommand{\bdis}{\begin{displaymath}}
\newcommand{\edis}{\end{displaymath}}
%------------------------------------------------------------------------------
\newcommand{\bit}{\begin{itemize}}
\newcommand{\eit}{\end{itemize}}
%------------------------------------------------------------------------------
\newcommand{\ben}{\begin{enumerate}}
\newcommand{\een}{\end{enumerate}}
%------------------------------------------------------------------------------
\newcommand{\nid}{\noindent}
\newcommand{\cl}{\centerline}
\newcommand{\nl}{\newline}
\newcommand{\ul}{\underline}
\newcommand{\dd}{\quad}
%------------------------------------------------------------------------------
%\newcommand{\Bbb}{{\cal I} \! \!}
\newcommand{\re}{{\cal I \! \!R}}
\newcommand{\co}{\,{ I \! \! \! \! C}}
\newcommand{\ze}{{\cal Z \! \! \! \!Z}}
\newcommand{\zp}{{\cal Z}(\beta)}
\newcommand{\zpn}{{\cal Z}(\beta, N)}
\newcommand{\zpc}{{\cal Z }_c ( \beta )}
\newcommand{\zpcn}{{\cal Z }_c ( \beta,N )}
\newcommand{\isn}{\Omega(v,N)}
\newcommand{\isnp}{\Omega^\prime(v,N)}
\newcommand{\imp}{\int {\cal D }[q]}
\newcommand{\imf}{\int {\cal D }[\phi]}
\newcommand{\ebvq}{\exp \{ - \beta \sum_i  V[q_i] \}}
\newcommand{\ebvf}{\exp \{ - \beta \int dx   V[\phi(x)] \}}
\newcommand{\ebvfd}{\exp \{ - \beta \sum_i   V[\phi] \}}
\newcommand{\zee}{{\cal Z \! \! \! \!Z}_2}
\newcommand{\boh}{\Longleftrightarrow}
\newcommand{\void}{\lpg \o \rpg}
\newcommand{\mm}{{\rho_{\mu}}}
\newcommand{\mc}{{\rho_{can}}}
\newcommand{\mgc}{{\rho_{gc}}}
\newcommand{\ham}{{\cal H}}
\newcommand{\Xc}{{\cal X}}
\newcommand{\Yc}{{\cal Y}}
\newcommand{\Tc}{{\cal T}}
\newcommand{\Mc}{{\cal M}}
\newcommand{\Cc}{{\cal C}}
\newcommand{\vb}{\bar{v}}
\newcommand{\ub}{\bar{u}}
\newcommand{\vbx}{{v \! \! \! \!  -}_{max}}
\newcommand{\vbi}{{v \! \! \! \!  -}_{min}}
\newcommand{\Kc}{{\cal K}}
\newcommand{\Rc}{{\cal R}}
\newcommand{\Mcv}{{\cal M}_v}
\newcommand{\Mcg}{{\cal M}^G_g}
\newcommand{\Mce}{{\cal M}^{\cal H}_E}
\newcommand{\Si}{\Sigma}
\newcommand{\Siv}{\Sigma_v}
\newcommand{\Sig}{\Sigma^G_g}
\newcommand{\Sie}{\Sigma^{\cal H}_E}
\newcommand{\nG}{\frac{\partial^\mu G}{\| \nabla G \|}}
\newcommand{\nGn}{\frac{\partial^\mu G}{\| \nabla G \|^2}}
\newcommand{\nH}{\frac{\partial^\mu {\cal H}}{\| \nabla {\cal H} \|}}
\newcommand{\nHn}{\frac{\partial^\mu {\cal H}}{\| \nabla {\cal H} \|^2}}
\newcommand{\ngH}{\| \nabla {\cal H} \|}
\newcommand{\ngV}{\| \nabla V \|}
\newcommand{\gV}{ \nabla V}
\newcommand{\de}{\partial}
\newcommand{\too}{\longrightarrow}
\newcommand{\impl}{\Longrightarrow}
\newcommand{\eb}{{\bf e}}
\newcommand{\lpt}{\left(}
\newcommand{\rpt}{\right)}
\newcommand{\lpg}{\{}
\newcommand{\rpg}{\}}
\newcommand{\lpq}{\left[}
\newcommand{\rpq}{\right]}
\newcommand{\e}{\mbox{\boldmath $e$}}
\newcommand{\x}{\mbox{\boldmath $x$}}
\newcommand{\y}{\mbox{\boldmath $y$}}
\newcommand{\vi}{\mbox{\boldmath $i$}}
\newcommand{\vj}{\mbox{\boldmath $j$}}
\newcommand{\qs}{\frac{dq}{ds}}
\newcommand{\q}{\mbox{\boldmath $q$}}
\newcommand{\p}{\mbox{\boldmath $p$}}
\newcommand{\Q}{\mbox{\boldmath $Q$}}
\newcommand{\bP}{\mbox{\boldmath $P$}}
\newcommand{\bal}{\mbox{\boldmath $\alpha$}}
\newcommand{\A}{{\bf A}}
\newcommand{\bS}{{\bf S}}
\newcommand{\cn}{\mbox{cn}}
\newcommand{\sn}{\mbox{sn}}
\newcommand{\lps}{\langle}
\newcommand{\rps}{\rangle}
\newcommand{\dyy}{\displaystyle}
\newcommand{\nnv}{\frac{\nabla}{\| \nabla V \|}}
\newcommand{\unv}{\frac{1}{\| \nabla V \|}}
\newcommand{\ps}{\underline{\psi}}
\newcommand{\psv}{\underline{\psi}(V)}
\newcommand{\psc}{\underline{\psi}(\chi)}
\newcommand{\psvi}{{\psi_i}(V)}
\newcommand{\psvj}{{\psi_j}(V)}
\newcommand{\psvk}{{\psi_k}(V)}
\newcommand{\psvr}{{\psi_r}(V)}
\newcommand{\psci}{{\psi_i}(\chi)}
\newcommand{\pscj}{{\psi_j}(\chi)}
\newcommand{\psck}{{\psi_k}(\chi)}
\newcommand{\pscr}{{\psi_r}(\chi)}
\newcommand{\psvd}{\underline{\psi}(V) \cdot}
\newcommand{\pscd}{\underline{\psi}(\chi) \cdot}
\newcommand{\dpsv}{\underline{ \cdot \psi}(V)}
\newcommand{\dpsc}{\underline{ \cdot \psi}(\chi)}
\newcommand{\emme}{\nabla \frac{\nabla V}{\| \nabla V \|}}
\newcommand{\dvnv}{\frac{\triangle V}{\| \nabla V \|^2}}
\newcommand{\dv}{\triangle V}
%\newcommand{\qed}{$\Box$}

%%%%%%%%%%%%%%%%%%%%%%%%%%%%%%%%%%%%%%%%%%%%%%%%%%%%%%%%%%%%%%%%%%%%%%%%%%%%%%%

\newtheorem{lemma}{Lemma}
\newtheorem{theorem}{Theorem}
\newtheorem{corollary}{Corollary}
\newtheorem{conjecture}{Conjecture}
\newtheorem{proposition}{Proposition}
\newtheorem{definition}{Definition}
\newtheorem{remark}{Remark}
\theoremstyle{definition}
\newtheorem*{pro}{Proof}

%%%%%%%%%%%%%%%%%%%%%%%%%%%%%%%%%%%%%%%%%%%%%%%%%%%%%%%%%%%%%%%%%%%%%%%%%%%%%%%

\title{Topology and Phase Transitions II. Theorem on a necessary relation}

\date{\today}

\author{Roberto Franzosi }

\affiliation{ Dipartimento di Fisica dell'Universit\`a di Firenze, Via G. Sansone 1, I-50019 Sesto Fiorentino, and C.N.R.-I.N.F.M., Italy }

\author{Marco Pettini\footnote{Corresponding author. e-mail: pettini@arcetri.astro.it,
Phone: +39-055-2752282, Fax: +39-055-220039.}  }

\affiliation{ Istituto Nazionale
di Astrofisica -- Osservatorio Astrofisico di Arcetri, Largo E. Fermi 5,
 50125 Firenze, Italy\\ and I.N.F.M., Unit\`a di Firenze, and I.N.F.N., Sezione di
Firenze}

\begin{abstract}
In this second paper, we prove a necessity Theorem about the
topological origin of phase transitions. We consider physical systems described by
smooth microscopic interaction potentials $V_N(q)$, among $N$ degrees of freedom,
and the associated family of configuration space submanifolds $\{ M_{v}\}_{v \in{\Bbb R}}$,
with $M_v=\{ q\in{\Bbb R}^N\vert V_N(q)\leq v\}$.
On the basis of an analytic relationship between a suitably weighed sum of the
Morse indexes of the manifolds
$\{ M_{v}\}_{v \in{\Bbb R}}$ and thermodynamic entropy,
the Theorem states that any possible unbound growth with $N$ of one of the following
derivatives of the configurational
entropy $S^{(-)}(v)=(1/N) \log \int_{M_v} d^Nq$, that is of
$\vert\partial^k S^{(-)}(v)/\partial v^k\vert$, for $k=3,4$, can be
entailed only by the weighed sum of Morse indexes.
 Since the unbound growth
with $N$ of one of these derivatives corresponds to the occurrence of a first or of a
second order
phase transition, and since the variation of the Morse indexes of a manifold
is in one-to-one correspondence with a change of its topology, the Main Theorem
of the present paper states that a phase transition {\it necessarily} stems from
a topological transition in configuration space.  The proof of the Theorem given in the present paper
cannot be done without Main Theorem of paper I.

\end{abstract}

\pacs{05.70.Fh; 05.20.-y; 02.40.-k}
\keywords{Statistical Mechanics, Phase Transitions, Topology}

\maketitle

%%%%%%%%%%%%%%%%%%%%%%%%%%%%%%%%%%%%%%%%%%%%%%%%%%%%%%%%%%%%%%%%%%%%%%%
%%%%%%%%%%%%%%%%%% BODY OF PAPER
%%%%%%%%%%%%%%%%%%%%%%%%%%%%%%%%%%%%%%%%%%%%%%%%%%%%%%%%%%%%%%%%%%%%%%%

\section{Introduction}
\label{introduction}
In Statistical Mechanics, a central task of the mathematical theory of
phase transitions has been to prove the loss of differentiability
of the pressure function -- or of other thermodynamic
functions -- with respect to temperature, or volume, or an external field.
The first rigorous results of this kind are the exact solution of
$2d$ Ising model due to Onsager \cite{onsager}, and the  Yang-Lee theorem
\cite{YLthm} showing that, despite the smoothness of the canonical and grand
canonical partition functions respectively, in the $N\rightarrow\infty$ limit
also piecewise
differentiability of pressure or other thermodynamic functions becomes
possible.

Another approach to the problem has considerably grown after the introduction
of the concept
of a Gibbs measure for infinite systems by Dobrushin, Lanford and Ruelle.
In this framework, the phenomenon of phase transition is seen as the
consequence of non-uniqueness of a Gibbs measure for a given type of
interaction among the particles of a system \cite{ruelleTD,georgii}.

Recently, it has been conjectured\cite{cccp,top1,top2,top3,physrep}
that the origin of the phase transitions
singularities could be attributed to suitable topology changes within the
family of equipotential hypersurfaces $\{ \Sigma_v=V_N^{-1}(v)\}_{v\in{\Bbb R}}$
of configuration space. These level sets of $V_N$ naturally foliate the
support
of the statistical measures (canonical or microcanonical) so that the
mentioned topology change would induce
a change of the measure itself at the transition point.
In a few particular cases, the truth of this {\it topological hypothesis}
has been given strong evidence: {\it i)} through the numerical computation
of the Euler characteristic for the $\{ \Sigma_v\}_{v\in{\Bbb R}}$ of a
two-dimensional lattice $\varphi^4$ model \cite{top2};
{\it ii)} through the exact analytic computation
of the Euler characteristic of
$\{ M_v = V_N^{-1}((-\infty,v])\}_{v\in{\Bbb R}}$
submanifolds of configuration space for two different models, the  mean-field $XY$
model\cite{xymf} and the $k$-trigonometric model \cite{ptrig}.

In the present paper we prove a necessity Theorem which implies that for a wide class of
potentials (good Morse functions), a first or a second order phase
transition can only be the consequence of a topology change of the submanifolds
$M_v$ of configuration space, and this appears to be the truly primitive and
deep mathematical origin of the phase transition phenomena, at least for the
mentioned class of potentials.

The Theorem is enunciated as follows:

\smallskip
\noindent{\bf Theorem } {\it Let $V_N(q_1,\dots,q_N): {\Bbb R}^N
\rightarrow {\Bbb R}$, be a smooth, non-singular, finite-range
potential. Denote by $M_v:= V_N^{-1}((-\infty,v])$, $v\in{\Bbb R}$, the
generic submanifold of configuration space bounded by $\Sigma_v$.
Let $\{ q_c^{(i)}\in{\Bbb R}^N\}_{i\in[1,{\cal N}(v)]}$ be the set of critical
points of the potential, that is s.t. $\nabla V_N(q)\vert_{q=q_c^{(i)}}=0$, and
${\cal N}(v)$ be the number of critical points up to the potential energy
value $v$. Let $\Gamma(q_c^{(i)},\varepsilon_0)$ be pseudo-cylindrical
neighborhoods of the critical points, and $\mu_i(M_v)$ be the Morse indexes
of $M_v$, then there exist real numbers $A(N,i,\varepsilon_0)$, $g_i$
and real smooth functions
$B(N,i,v,\varepsilon_0)$ such that the following equation for the
microcanonical configurational entropy $S_N^{(-)}(v)$ holds

\begin{eqnarray}
S_N^{(-)}(v) &=&\frac{1}{N} \log \left[ \int_{M_v
\setminus\bigcup_{i=1}^{{\cal N}(v)}
\Gamma(q^{(i)}_c,\varepsilon_0)}\ d^Nq + \sum_{i=0}^N
A(N,i,\varepsilon_0 ) \ g_i\ \mu_i (M_{v-\varepsilon_0})\right. \nonumber\\
&+&\left. \sum_{n=1}^{{\cal
N}_{cp}^{\nu(v)+1}}B(N,i(n),v-v_c^{\nu(v)},\varepsilon_0 )
  \right] \ ,\nonumber
\end{eqnarray}
(details and appropriate definitions are given in Section \ref{sec2}),
and an unbound growth with $N$ of one of the derivatives
$\vert\partial^k S^{(-)}(v)/\partial v^k\vert$, for $k=3,4$, and thus the
occurrence of a first or of a second order phase transition respectively, can
be entailed only by the topological term $\sum_{i=0}^N A(N,i,\varepsilon_0 )\
g_i\ \mu_i (M_{v-\varepsilon_0})+ \sum_{n=1}^{{\cal
N}_{cp}^{\nu(v)+1}}B(N,i(n),v-v_c^{\nu(v)},\varepsilon_0 )$. }
\smallskip

The above given expression for the entropy stems from a decomposition of the volume
of a generic
submanifold $M_v$ into two parts: the volume of the disjoint union of suitably defined
neighborhoods of the
critical points of the potential $V_N(q)$, and the volume of its complement.
The latter is represented by the first term in square parentheses, and, after Main
Theorem of paper I, it cannot entail unbounded growth with $N$ of up to the fourth derivative
of the entropy. Only the second and third terms in square parentheses could do it.
These two terms, representing the volume of the neighborhoods of critical points, are of
topological meaning.

Thus the proof of the present Theorem {\it crucially} relies on Main Theorem of paper I
which is a first step toward proving the necessity of topology changes
of configuration space submanifolds (either the level sets $\Sigma_v$ of the potential
function, or the manifolds $M_v$ bounded by them) for the
occurrence of phase transitions \cite{paperI,pirl}.
Main Theorem of paper I is based on the assumption of the
existence -- at any number $N$ of degrees of freedom -- of an energy density
interval $[\vb_0,\vb_1]$ free of critical values. Under this assumption, in paper I,
we proved the uniform convergence in the thermodynamic limit of configurational entropy
in the class of three times differentiable functions. However, since in general it is a
very hard task to locate all the critical points of a given potential function,
it is also very hard to ascertain whether Main Theorem of paper I applies to it or not.
To overcome this limitation, in the present paper we prove a new Theorem, based on
a less restrictive assumption (allowing the existence of critical points),
which we could not prove without the Main Theorem
of paper I.

As already remarked above, the necessity Theorem proved in the present paper applies to
a very broad class of systems: those described by finite range potentials which are good
Morse functions. In fact, checking whether
a given potential is a good Morse function or not is not difficult and amounts
to control whether the potential is smooth, bounded below and whether its
Hessian is non-degenerate, that is free of vanishing eigenvalues; degeneracy
typically occurs in presence of continuous symmetries and can be removed by
arbitrarily small and standard perturbations \cite{MorseCairns}
which do not alter neither the
microscopic dynamics produced by the gradients of the potential nor the
thermodynamics.

%============================================================================

\section{Basic definitions}
\label{sec2}

For a physical system ${\cal S}$ of $n$ particles confined in a
bounded subset  $\Lambda^d$ of ${\Bbb R}^d$, $d=1,2,3$,  and
interacting through a real valued
potential function $V_N$ defined on
$(\Lambda^d)^{\times n}$, with $N=n d$, the {\it configurational
microcanonical volume}
$\Omega(v,N)$ is defined for any value $v$ of the potential $V_N$ as
\begin{equation}
\Omega(v,N) = \int_{(\Lambda^d)^{\times n}} dq_1\dots dq_N\
\delta[V_N(q_1,\dots, q_N) - v]
= \int_{\Sigma_v}\ \frac{d\sigma}{\Vert \nabla V_N\Vert}~,
\label{mi_volume}
\end{equation}
where $d\sigma$ is a surface element of $\Si_v:=V_N^{-1}(v)$; in what
follows $\Omega(v,N)$ is also called {\it structure integral}. The norm
$\Vert \nabla V_N\Vert$ is defined as $\Vert \nabla V_N\Vert =[\sum_{i=1}^N
(\partial_{q_i}V_N)^2]^{1/2}$.

Now we can define the configurational thermodynamic functions to be used in
this paper.

Henceforth, according to the need for explicit reference to the
$N$-dependence of $V$, we shall use both $V$ and $V_N$ to denote the potential.

\begin{definition}
Using the notation  $\vb = v /N$ for the value of the potential energy per
particle, we introduce the following functions:

- {\em Configurational microcanonical entropy, relative
  to $\Sigma_v$}.
  For any $N\in{\Bbb N}$ and
       $\vb\in{\Bbb R}$,
    \begin{eqnarray}
     S_N(\vb)\equiv S_N(\vb;V_N)
          =\frac{1}{N}
      \log{\Omega(N \vb, N)} \, .
    \nonumber
    \end{eqnarray}

- {\em Configurational microcanonical entropy,
relative to the volume bounded by $\Sigma_v$}.
For any $N\in{\Bbb N}$ and
        $\vb\in{\Bbb R}$,
        \begin{eqnarray}
      S^{(-)}_N(\vb) \equiv S^{(-)}_N(\vb;V_N)
          =\frac{1}{N} \log{ M (N \vb, N)} \,
    \nonumber
    \end{eqnarray}
where
\begin{equation}
M (v,N) = \int_{(\Lambda^d)^{\times n}} dq_1\dots dq_N\
\Theta [V_N(q_1,\dots, q_N) - v]
=\int_0^v d\eta \  \int_{\Sigma_\eta}\ \frac{d\sigma}{\Vert \nabla V_N\Vert}~,
\label{pallaM}
\end{equation}
with $\Theta[\cdot]$ the Heaviside step function; $M(v,N)$ is the
codimension-0 subset of configuration space enclosed by the equipotential
hypersurface $\Sigma_v$. The representation of $M(v,N)$ given in the r.h.s.
stems from the already mentioned co-area formula in \cite{federer}.
\end{definition}

\begin{definition}[First and second order phase transitions]
\label{PTs}
We say that a physical system ${\cal S}$ undergoes a phase transition if
there exists a thermodynamic function which -- in the thermodynamic limit
($N\rightarrow\infty$ and ${ vol}(\Lambda^d )/N={ const}$) -- is
only piecewise analytic.
In particular, if the second-order derivative of the entropy
$S^{(-)}_\infty(\vb )$ is discontinuous at some point ${\vb}_c$, then we say
that a {\it first-order} phase transition occurs.
If the third-order derivative of the entropy $S^{(-)}_\infty(\vb )$ is
discontinuous at some point ${\vb}_c$, then we say
that a {\it second-order} phase transition occurs.
These definitions stem from the standard definitions of first and second order
phase transitions as due to a discontinuity of the first or second derivatives
of the Helmoltz free energy, respectively, and from the existing
relationship -- through a Legendre transform -- between the Helmoltz free
energy and the entropy (see Definition 1 in paper I).
\end{definition}

\begin{definition}[Standard potential, fluid case]
\label{st-pot}
We say that an $N$ degrees of freedom potential $V_N$ is a
{\emph standard potential} for a fluid if it is of the form
\begin{eqnarray}
        V_N:& & {\cal B}_N\subset{\Bbb R}^N \rightarrow {\Bbb R}\nonumber\\
    V_N(q) &=& \sum_{i\neq j=1}^n
    \Psi (\| \vec{q}_i - \vec{q}_j  \|)
% +\sum_{i =1}^n \Phi ( \vec{q}_i )
    + \sum_{i =1}^n U_\Lambda ( \vec{q}_i )\,
\end{eqnarray}
where ${\cal B}_N$ is a compact subset of ${\Bbb R}^N$, $N=nd$,
$\Psi$ is a real valued function of one variable such that additivity
holds, and where $U_\Lambda$ is any smoothed potential barrier to confine the
particles in a finite volume $\Lambda$, that is
\[
U_\Lambda ( \vec{q} )=\left\{ \begin{array}{cc}
0&~if~\vec{q}\in\Lambda^\prime \\
+\infty&~if~\vec{q}\in\Lambda^c,~ complement~ in~ {\Bbb R}^N \\
{\cal C}^\infty &function~for~\vec{q}\in\Lambda\setminus\Lambda^\prime
\end{array} \right.
\]
where $\Lambda^\prime\subset\Lambda$ and $\Lambda^\prime$ arbitrarily close
to $\Lambda\subset{\Bbb R}^N$, closed and bounded. $U_\Lambda$ is a confining
potential in a limited spatial volume with the additional property that
given two limited $d$-dimensional regions of space, $\Lambda_1$ and
$\Lambda_2$, having in common a $d-1$-dimensional boundary,
$U_{\Lambda_1}+U_{\Lambda_2}=U_{\Lambda_1\cup\Lambda_2}$.
By additivity we mean what follows. Consider two systems ${\cal S}_1$ and
${\cal S}_2$, having $N_1=n_1 d$ and $N_2=n_2 d$ degrees of freedom, occuping
volumes
$\Lambda_1^d$ and $\Lambda_2^d$, having potential energies $v_1$  and $v_2$,
for any $(q_1, \ldots ,q_{N_1})\in (\Lambda_1^d)^{\times n_1}$ such that
$V_{N_1}(q_1, \ldots ,q_{N_1}) =v_1$,
for any $(q_{N_1+1}, \ldots ,q_{N_1+N_2})\in (\Lambda_2^d)^{\times n_2}$
such that $V_{N_2}(q_{N_1+1},\ldots ,q_{N_1+N_2}) = v_2$,
for  $(q_1, \ldots ,q_{N_1+N_2})\in (\Lambda_1^d)^{\times n_1}\times
(\Lambda_2^d)^{\times n_2}$ let $V_N(q_1,\ldots ,q_{N_1+N_2}) = v$ be
the potential energy $v$ of the compound system
${\cal S}={\cal S}_1+{\cal S}_2$
which occupies the volume  $\Lambda^d = \Lambda_1^d\cup\Lambda_2^d$ and
contains $N=N_1+N_2$ degrees of freedom. If
    \begin{equation}
         v(N_1+N_2,\Lambda_1^d\cup\Lambda_2^d) = v_1(N_1,\Lambda_1^d)
      + v_2(N_2,\Lambda_2^d) + v^\prime (N_1,N_2,\Lambda_1^d,\Lambda_2^d)
        \label{V_somma}
    \end{equation}
where $v^\prime$ stands for the interaction energy between ${\cal S}_1$ and
${\cal S}_2$, and if $v^\prime /v_1\rightarrow 0$ and
$v^\prime /v_2\rightarrow 0$ for $N\rightarrow\infty$ then $V_N$ is additive.
Moreover, at short distances $\Psi$ must be a repulsive
potential so as to prevent the concentration of an arbitrary number of particles
within small, finite volumes of any given size.
\end{definition}

\begin{definition}[Standard potential, lattice case]
\label{st-pot1}
We say that an $N$ degrees of freedom potential $V_N$ is a
{\emph standard potential} for a lattice if it is of the form
\begin{eqnarray}
        V_N:& & {\cal B}_N\subset{\Bbb R}^N \rightarrow {\Bbb R}\nonumber\\
    V_N(q) &=& \sum_{{\underline i},{\underline j}\in{\cal I}\subset{\Bbb N}^d}
 C_{{\underline i} {\underline j}}
    \Psi (\| \vec{q}_{\underline i} - \vec{q}_{\underline j}  \|) +
    \sum_{{\underline i}\in{\cal I}\subset{\Bbb N}^d}
     \Phi ( \vec{q}_{\underline i} )
\end{eqnarray}
where ${\cal B}_N$ is a compact subset of ${\Bbb R}^N$. Denoting by
$a_1,\dots,a_d$ the lattice spacings, if ${\underline i}\in{\Bbb N}^d$, then
$(i_1a_1,\dots,i_da_d)\in\Lambda^d$. We denote by $m$ the number of lattice
sites in each spatial direction, by $n=m^d$ the total number of lattice sites,
by $D$ the number of degrees of freedom on each site. Thus
$\vec{q}_{\underline i}\in{\Bbb R}^D$ for any ${\underline i}$. The total
number of degrees of freedom is $N=m^dD$. Having two systems made of $N=m^dD$
degrees of freedom, whose site indexes
${\underline i}^{(1)}$ and ${\underline i}^{(2)}$ run over
$1\leq {i}_1^{(1)},\dots, { i}_d^{(1)}\leq m$, and
$1\leq {i}_1^{(2)},\dots, { i}_d^{(2)}\leq m$, after
gluing together the two systems through a common $d-1$ dimensional boundary
the new system has indexes ${\underline i}$ running over, for example,
$1\leq {i}_1\leq 2m$ and $1\leq {i}_2,\dots,
{i}_d\leq m$. If
\begin{equation}
        v(N+N,\Lambda_1^d\cup\Lambda_2^d) = v_1(N,\Lambda_1^d)
   + v_2(N,\Lambda_2^d) + v^\prime (N,N,\Lambda_1^d,\Lambda_2^d)
 \label{V_summa}
 \end{equation}
where $v^\prime$ stands for the interaction energy between the two systems and
if $v^\prime /v_1\rightarrow 0$ and
$v^\prime /v_2\rightarrow 0$ for $N\rightarrow\infty$ then $V_N$ is additive.
\end{definition}

\begin{definition}[Short-range potential]
\label{sr-pot}
In defining a short-range potential, a distinction has to be made between
lattice systems and fluid systems.
Given a standard potential $V_N$ on a lattice, we say that it is a short-range
potential if the coefficients $C_{{\underline i}{\underline j}}$
are such that for any ${\underline i},{\underline j}\in{\cal I}\subset{\Bbb N}^d$,
$C_{{\underline i}{\underline j}}=0$
iff $\vert {\underline i} - {\underline j}\vert > c$, with $c$ is
definitively constant for
$N\rightarrow\infty$.

Given a standard potential $V_N$ for a fluid system, we say that it is a
short-range potential if there exist $R_0>0$ and $\epsilon >0$ such that
for $\Vert{\bf q}\Vert >R_0$ it is
$\vert\Psi (\Vert{\bf q}\Vert )\vert < \Vert{\bf q}\Vert^{-(d+\epsilon)}$,
where $d=1,2,3$ is the spatial dimension.
\end{definition}

\begin{definition}[Stable potential]
\label{stab-pot}
We say that a potential $V_N$ is stable \cite{ruelle} if there exists
$B\geq 0$ such that
\begin{equation}
V_N(q_1,\dots,q_N)\geq -N B
\end{equation}
for any $N > 0$ and $(q_1,\dots,q_N)\in (\Lambda^d)^{\times n}$, or for
$\vec{q}_{\underline i}\in{\Bbb R}^D$,
${\underline i}\in{\cal I}\subset{\Bbb N}^d$, $N=m^dD$, for lattices.
\end{definition}

\begin{definition}[Confining potential]
\label{conf-pot}
With the above definitions of standard potentials $V_N$, in the fluid case the
potential is said to be confining in the sense that it contains $U_\Lambda$
which constrains the
particles in a finite spatial volume, and in the lattice case the potential
$V_N$ contains an on-site potential such that -- at finite energy --
$\Vert \vec{q}_{\underline i}\Vert$ is constrained in compact set of values.
%If $\Lambda^d \equiv{\Bbb R}^d$, a standard potential $V$
%is said to be a confining potential when $V(q)\rightarrow\infty$
%whenever $\Vert  \vec{q}_i\Vert \rightarrow\infty$ or
%$\Vert  q_i -q_j\Vert \rightarrow\infty$.
%This  means that at finite potential energy no
%particle can escape arbitrarily far away.
\end{definition}

\begin{remark}[Compactness of equipotential hypersurfaces]
From the previous definition it follows that, for a confining potential, the
equipotential hypersurfaces $\Sigma_v$ are compact (because they are closed
by definition and bounded in view of particle confinement).
\end{remark}

%%%%%%%%%%%%%%%%%%%%%%%%%%%%%%%%%%%%%%%%%%%%%%%%%%%%%%%%%%%%%%%%%%%%%%%%%%%%%
\noindent In view of formulating and proving the Main Theorem of the
present paper, we have to
define some neighborhoods, that we call ``pseudo-cylindrical'', of critical
points of a potential function $V_N$.
Before defining these pseudo-cylindrical neighborhoods of critical
points, let us remember the following basic result in Morse
theory.

\smallskip
\noindent{\bf Theorem.\ } {\it
%\begin{theorem}
Let $f$ be a smooth real valued function on a compact finite
dimensional manifold $M$. Let $a<b$ and suppose that the set
\begin{equation}
f^{-1}([a,b])\equiv {\cal M} =\{ x\in M\vert a\leq f(x) \leq b\}
\end{equation}
is compact and contains no critical points of $f$, that is
$\Vert\nabla f\Vert\geq C>0$ with $C$ a constant. Let $y\in
(a,b)$. Then there exists a diffeomorphism
\begin{equation}
\sigma :(a,b) \times f^{-1}(y) \rightarrow f^{-1}[(a,b)]~~ {\rm
by}~~ (v,x)\rightarrowtail \sigma (v,x).
\end{equation}   }
%\end{theorem}
\smallskip
\noindent{\bf Corollary.\ } {\it
%\begin{corollary}
The manifolds $f^{-1}(y)$, $a< y< b$, are all diffeomorphic.}
%\end{corollary}

This result is based on the existence of a one-parameter group of
diffeomorphism
\begin{equation}
\sigma_v : M \rightarrow {\cal M} ~~{\rm by}~~ x\rightarrowtail
\sigma (v,x)
\end{equation}
associated with the vector field $X =\nabla f(x)/\Vert\nabla
f(x)\Vert^2$ with $v\rightarrowtail\sigma (v,x)$ a solution of the
differential equation on $M$
\begin{equation}
\frac{d\sigma(v,x)}{dv}=\frac{\nabla f[\sigma(v,x)]}{\Vert\nabla
f[\sigma(v,x)]\Vert^2}\ , ~\sigma(0,x)=x.
\end{equation}
$\sigma(v,x)$ is defined for all $v\in{\Bbb R}$ and $x\in M$.
Details can be found in standard references as
\cite{palais,hirsch,milnor}.

Applied to the configuration space $M$, if the function $f$ is
identified with the potential $V_N$, then in the absence of critical
points of $V$ in the interval $(v_0,v_1)$ the hypersurfaces
$\Sigma_{v}=V_N^{-1}(v)$, $v\in (v_0,v_1)$, are all
diffeomorphic.

\begin{definition}[Pseudo-cylindrical neighborhoods]
Let $\Sigma_{v_c}$ be a critical level set of $\ V_N$, that is a level
set containing at least one critical point of $\ V_N$. Around any
critical point $q_c^{(i)}$, consider the set of points
$\gamma(q_c^{(i)},\rho ; v_c )\subset \Sigma_{v_c}$ at a distance
equal to $\rho >0$ from $q_c^{(i)}$, that is $q\in\gamma
(q_c^{(i)},\rho ; v_c)$ $\Rightarrow$ $d(q -q_c^{(i)}) =\rho$,
where $d(\cdot ,\cdot)$ is the distance measured through the
metric induced on $\Sigma_{v_c}$ by the euclidean metric of the
immersion space, and
$\rho$ is such that $\rho <\frac{1}{2}\min_{i,j} d(q_c^{(i)}-q_c^{(j)})$,
$i,j$ label all the critical points on
the given critical level set. Moreover, set the thickness of all
the pseudo-cylinders equal to $\varepsilon_0 = \min_{j\in{\Bbb
N}}(v_c^{j+1} - v_c^j)$. After Sard Theorem, both $\rho$ and
$\varepsilon_0$ are finite because, at finite dimension, there is
a finite number of isolated critical points and, consequently, a
finite number of critical values. We define a pseudo-cylindrical
neighborhood $\Gamma (q_c^{(i)},\varepsilon_0)\subset M$ of
$q_c^{(i)}$ as the open subset of $M$ bounded by the following set of
points. By mapping $\gamma(q_c^{(i)},\rho ; v_c )$ from
$\Sigma_{v_c}$ to $\Sigma_{(v_c +{\varepsilon}_0)}$, and from
$\Sigma_{v_c}$ to $\Sigma_{(v_c -{\varepsilon}_0)}$, through
the flow generated by the vector field $X =\nabla V_N(q)/\Vert\nabla
V_N(q)\Vert^2$, we obtain the walls of
$\Gamma(q_c^{(i)},\varepsilon_0)$, which are transverse to the
$\Sigma_{v}$, and then we close the neighborhood with the pieces of
$\Sigma_{(v_c +{\varepsilon}_0)}$ and
$\Sigma_{(v_c -{\varepsilon}_0)}$ bounded by the images
$\gamma(q_c^{(i)},\rho ;(v_c +{\varepsilon}_0 ))$ and
$\gamma(q_c^{(i)},\rho ;(v_c -{\varepsilon}_0) )$ of
$\gamma(q_c^{(i)},\rho ;v_c)$ through $\sigma(v,x)$, respectively.
\end{definition}

%%%%%%%%%%%%%%%%%%%%%%%%%%%%%%%%%%%%%%%%%%%%%%%%%%%%%%%%%%%%%%%%%%%%%%%%%%%%%

\begin{lemma}[Generalization of Corollary 1, paper I]
Let $V_N$ be a standard, smooth, confining, short-range potential bounded from
below (Definitions \ref{st-pot}--\ref{conf-pot})
\[
  V_N:  {\cal B}_N\subset{\Bbb R}^N \rightarrow {\Bbb R}
\]
with $V_N$ given by Definition \ref{st-pot} (fluid case), or by Definition
\ref{st-pot1} (lattice case).

Let $\{\Sigma_v\}_{v\in{\Bbb R}}$ be the family of $N-1$ dimensional
hypersurfaces $\Sigma_v := V_N^{-1}(v)$, $v\in{\Bbb R}$, of ${\Bbb R}^N$.
Let $\{M_v\}_{v\in{\Bbb R}}$ be the family of $N$ dimensional
subsets $M_v := V_N^{-1}((-\infty,v])$, $v\in{\Bbb R}$, of ${\Bbb R}^N$.
Let $\{\overline M_v\}_{v\in{\Bbb R}}$ be the family of $N$ dimensional
subsets $\overline M_v:=M_v\setminus\bigcup_{i=1}^{{\cal N}(v)}
\Gamma(q^{(i)}_c,\varepsilon)$,  $v\in{\Bbb R}$, of ${\Bbb R}^N$, where
$\Gamma(q^{(i)}_c,\varepsilon)$ are  the pseudo-cylindrical neighborhoods of
the critical points $q^{(i)}_c$ of $V_N(q)$ contained in $M_v$ and
${\cal N}(v)$ is the number of critical points in $M_v$.
Let $\{\overline \Sigma_v\}_{v\in{\Bbb R}}$ be the family of $N-1$ dimensional
subsets of ${\Bbb R}^N$ defined as $\overline\Sigma_v := \Sigma_v\setminus
\bigcup_{i=1}^{{\cal N}(v)}[\Gamma(q^{(i)}_c,\varepsilon)\cap \Sigma_v]$.

\noindent Let $\vb_0=v_0/N,\vb_1=v_1/N \in {\Bbb R}$, $\vb_0 < \vb_1$ and let
$\vb_c =v_c/N$ be the only critical value of $V_N$ in the interval
$I_{\vb}=[\vb_0 ,\vb_1]$, and let $\Gamma^\star(q^{(i)}_c,\varepsilon^\star)$,
with $q_c^{(i)}\in V^{-1}_N(v_c)$ and $\varepsilon^\star$ such that
$\varepsilon^\star > \max (v_1-v_c,v_c-v_0)$.
The following two statements hold:

a) for any $\vb,\vb' \in[\vb_0,\vb_1]$  it is
\[
\overline\Sigma_{N\vb}~is~ {\cal C}^\infty-{\rm diffeomorphic}~~to~
\overline\Sigma_{N\vb'};
\]

b) putting ${\overline M}(v,N)= {\rm vol}(\overline M_v)$, the
quantities $[d{\overline M}(v,N)/dv]/{\overline M}(v,N)$ and $(d^k/dv^k)
\{[d{\overline M}(v,N)/dv]/{\overline M}(v,N)\}$, $k=1,2,3$, are uniformly
bounded in $N$ in the interval $[\vb_0,\vb_1]$.
\label{Lm1}
\end{lemma}

\begin{pro}
\label{proofLm1}

For what concerns point {\it (a)}, we note that the flow associated with the
${\cal C}^\infty$ vector field $X =\nabla V_N(q)/\Vert\nabla V_N(q)\Vert^2$
is well defined at
any point $q\in \overline M_{v_1}\setminus \overline M_{v_0}$. Thus the set
$\overline M_{v_1}\setminus \overline M_{v_0}$ is diffeomorphic to the
non-critical neck $\partial \overline M_{v_0}\times [v_0,v_1]$. Then,
after the ``non-critical neck theorem''
\cite{palais}, for any $\vb,\vb' \in I_{\vb}=[\vb_0,\vb_1]$ it is
$\overline\Si_{N\vb}\approx \overline\Si_{N\vb'}$. Incidentally, this entails
also $\overline M_{N\vb}\approx \overline M_{N\vb'}$ for any
$\vb,\vb' \in [\vb_0,\vb_1]$.

Now let us consider point {\it (b)}. Define $\overline S_N(\vb)=\frac{1}{N}
\log [\overline\Omega (N\vb, N)]$, where $\overline\Omega (N\vb, N)=
{\rm vol}(\overline\Sigma_{\vb N})]$, having proved the statement {\it (a)},
we can apply Lemma 4 of paper I which entails that
    \bena
    \sup_{N,\vb\in I_{\vb}} \left\vert\overline S_N({\vb})\right\vert
        < \infty~~~{\it and}~~~
    \sup_{N,\vb\in I_{\vb}} \left\vert \frac{\de^k\overline
    S_N}{\de {\vb}^k}({\vb})\right\vert < \infty,~~k=1,2,3,4. \nonumber
    \eena
Whence, after Lemma 3 of paper I, it follows
$\overline S_\infty(\vb)=\lim_{N\to\infty}\overline S_N({\vb})
\in{\cal C}^3(I_{\vb})$.

The next step is to prove that also $\overline S_\infty^{(-)}(\vb)\in{\cal C}^3
(I_{\vb})$, where
\[
\overline S_\infty^{(-)}(\vb):=\lim_{N\to\infty}\overline S_N^{(-)}
(\vb)=\lim_{N\to\infty}\frac{1}{N}\log [{\rm vol}
(\overline M_{\vb N})],
\]
because, after Lemmas 3 and 4 of paper I, this entails the truth
of statement {\it (b)}.
Let us begin by considering the microcanonical configurational inverse
temperature. From its definition
$\beta_N(\vb)={\partial S_N^{(-)}}/{\partial \vb}$ one obtains
$\beta_N(\vb)=\Omega (N \vb, N)/M (N \vb, N)$. The function
$\beta_N(\vb)$ is well known to be intensive and well defined also in the
thermodynamic limit, at least for extensive potential energy functions.
Then we work out a representation of $\beta_N(\vb)$ in the form of a
microcanonical average of a suitable function. To this purpose we derive
$\Omega (N \vb, N)$ with respect to $v$ by means of Federer's derivation
formula \cite{federer,laurence}, and then we integrate it. Federer's derivation
formula (see Lemma 5 of paper I) states that
   \bena
\frac{d}{dv}\isn = \int_{\Si_v}\ngV \ A \lpt \unv \rpt \frac{d \sigma}
               {\ngV} \dd ,
   \label{derfed}
   \eena
where $A$ stands for the operator
\[
A(\bullet ) =\nabla \lpt \frac{\nabla V}{\| \nabla V \|}\ \bullet \rpt \frac{1}
{\| \nabla V \|}~.
\]
Then we can write
\begin{eqnarray}
\Omega (N \vb, N)&=&\int_0^v\ d\eta \int_{\Si_\eta}\ngV \ A \lpt \unv \rpt
\frac{d \sigma} {\ngV} \dd\nonumber\\
&=&\int_{M_v}\ d\mu\ \ngV \ A \lpt \unv \rpt
\label{coarea}
\end{eqnarray}
where $d\mu =d^Nq$, so that we finally obtain
\begin{equation}
\beta_N(\vb)=\left[\int_{M_v}\ d\mu\right]^{-1}\int_{M_v}\ d\mu\ \ngV \ A
\lpt \unv \rpt =\left\langle \ngV \ A \lpt \unv \rpt\right\rangle_{M_v}
\label{Taver}
\end{equation}
which holds for $\vb\in[0,\vb_0)$.
An important remark is in order. We have used Federer's derivation
formula apparently ignoring that it applies in the absence of critical
points of the potential function. However, if the potential $V$ is a good
Morse function (not a very restrictive condition at all) we know,
after Sard theorem \cite{palais}, that the ensemble of critical
values, here of the potential, is a point set. Therefore, any finite
interval of values of the potential is the union of a finite number of open
intervals where no critical value is present, and correspondingly no critical
point on the $\{\Sigma_v\}$ exists. On all these open sets, free of
critical points, Federer's derivation formula can be legally
applied. Moreover, the results found by applying Federer's formula on each
open interval free of critical values of $V$ can be regularly glued together
because of the existence of the thermodynamic limit of $\beta_N(\vb)$.

Let us now consider $\overline\Omega (N \vb, N)$ for $\vb\in[\vb_0,\vb_1]$.
As all the hypersurfaces $\overline\Si_{\vb}$ labeled by $\vb\in[\vb_0,\vb_1]$
are diffeomorphic, we can use Federer's derivation formula to obtain an
expression for $\overline\Omega (N \vb, N)$ similar to that given in
Eq.(\ref{coarea}) for $\Omega (N \vb, N)$, that is
\begin{eqnarray}
\overline\Omega (N \vb, N)&=&\int_{v_0}^v\ d\eta \int_{\overline\Si_\eta}
\ngV \ A \lpt \unv \rpt \frac{d \sigma} {\ngV} + \Omega (N \vb_0, N)
\nonumber\\
&=&\int_{\overline M_v\setminus M_{v_0}}\ d\mu\ \ngV \ A \lpt \unv \rpt +
\int_{M_{v_0}}\ d\mu\ \ngV \ A \lpt \unv \rpt\nonumber\\
&=& \int_{M_v\setminus \Gamma^\star}\ d\mu\ \ngV \ A \lpt \unv \rpt
\label{omegabarra}
\end{eqnarray}
where $\Gamma^\star$ stands for the union of all the pseudo-cylindrical
neighborhoods of the critical points of $V$ in the interval $[\vb_0,\vb_1]$.
Then we consider the restriction $\overline\beta_N(\vb)$ of the function
$\beta_N(\vb)$ to the subset $M_v\setminus \Gamma^\star$;
from
\begin{equation}
\overline\beta_N(\vb)=\frac{\overline\Omega (N \vb, N)}{\overline M (N \vb, N)}
\label{betabarra}
\end{equation}
we get
\begin{eqnarray}
\overline\beta_N(\vb)&=&\left[\int_{M_v\setminus\Gamma^\star}\ d\mu
\right]^{-1}\int_{M_v\setminus\Gamma^\star}\ d\mu\ \ngV \ A
\lpt \unv \rpt \nonumber\\
&=&\left\langle \ngV \ A \lpt \unv \rpt
\right\rangle_{M_v\setminus\Gamma^\star}.
\label{Taverbar}
\end{eqnarray}
By comparing Eq.(\ref{Taver}) with Eq.(\ref{Taverbar}), we see that also
$\overline\beta_N(\vb)$ has to be intensive up to the $N\to\infty$ limit,
like $\beta_N(\vb)$. In fact the excision of the set $\Gamma^\star$ out of
$M_v$, no matter how the measure of $\Gamma^\star$ depends on $N$, cannot
change the intensive character of $\overline\beta_N(\vb)$. The relationship
among $\overline S_N(\vb )$, $\overline S^{(-)}_N(\vb )$ and
$\overline\beta_N(\vb)$ is given by the logarithm of both
sides of (\ref{betabarra})
\[
%\begin{equation}
\frac{1}{N}\log \overline\Omega(\vb N,N) =
\frac{1}{N}\log \overline M (\vb N,N) +
\frac{1}{N}\log \overline\beta_N(\vb)\ .
%\label{equi}
%\end{equation}
\]
whence, using $\lim_{N\to\infty}\frac{1}{N}\log \overline\beta_N(\vb) = 0$,
we obtain $\overline S_\infty^{(-)}(\vb)=\overline S_\infty(\vb)$ and
thus $\overline S_\infty^{(-)}(\vb)\in{\cal C}^3(I_{\vb})$.

Finally, $\overline S_\infty^{(-)}(\vb)\in{\cal C}^3(I_{\vb})$ entails
    \bena
    \sup_{N,\vb\in I_{\vb}} \left\vert\overline S^{(-)}_N({\vb})\right\vert
        < \infty~~~{\it and}~~~
    \sup_{N,\vb\in I_{\vb}} \left\vert \frac{\de^k\overline
    S^{(-)}_N}{\de {\vb}^k}({\vb})\right\vert < \infty,~~k=1,2,3,4. \nonumber
    \eena
so that, resorting to Lemma 3 of paper I, the truth of statement {\it (b)}
follows. \qed
\end{pro}

%$%%%%%%%%%%%%%%%%%%%%%%%%%%%%%%%%%%%%%%%%%%%%%%%%%%%%%%%%%%%%%%%%%%%%%%%%%%%%%
\section{Main Theorem}
\label{mainthm2}
In this Section we prove the following

\begin{theorem}[Entropy and Topology] Let $V_N(q_1,\dots,q_N): {\Bbb R}^N
\rightarrow {\Bbb R}$, be a smooth, non-singular, finite-range
potential. Denote by $M_v:= V_N^{-1}((-\infty,v])$, $v\in{\Bbb R}$, the
generic submanifold of configuration space bounded by $\Sigma_v$.

Let $\{ q_c^{(i)}\in{\Bbb R}^N\}_{i\in[1,{\cal N}(v)]}$ be the set of critical
points of the potential, that is s.t. $\nabla V_N(q)\vert_{q=q_c^{(i)}}=0$, and
${\cal N}(v)$ be the number of critical points up to the potential energy
value $v$. Denote by $\vb =v/N$ the potential energy density, and assume that
for any given interval $[\vb_0,\vb_1]$ the number of critical values
$\vb_c^j$ contained in it is at most a linearly growing function of $N$.
Let $\Gamma(q_c^{(i)},\varepsilon_0)$ be the pseudo-cylindrical
neighborhood of the critical point $q_c^{(i)}$, and $\mu_i(M_v)$ be the Morse
indexes of $M_v$, then there exist real numbers $A(N,i,\varepsilon_0)$, $g_i$
and real smooth functions
$v\rightarrowtail B(N,i,v,\varepsilon_0)$ such that the following equation
for the microcanonical configurational entropy $S_N^{(-)}(v)$ holds
\begin{eqnarray}
S_N^{(-)}(v) &=&\frac{1}{N} \log \left[ \int_{M_v
\setminus\bigcup_{i=1}^{{\cal N}(v)}
\Gamma(q^{(i)}_c,\varepsilon_0)}\ d^Nq + \sum_{i=0}^N
A(N,i,\varepsilon_0 ) \ g_i \ \mu_i (M_{v-\varepsilon_0})\right. \nonumber\\
&+&\left. \sum_{n=1}^{{\cal
N}_{cp}^{\nu(v)}}B(N,i(n),v-v_c^{\nu(v)},\varepsilon_0 )
  \right] \ ,\label{meaning}
\end{eqnarray}
where $\nu(v)=\max \{ j\vert v_c^j\leq v\}$,
and an unbound growth with $N$ of one of the derivatives
$\vert\partial^k S^{(-)}(v)/\partial v^k\vert$, for $k=3,4$, and thus the
occurrence of a first or of a second order phase transition respectively, can
be entailed only by the topological term $\sum_{i=0}^N A(N,i,\varepsilon_0 )\
g_i \ \mu_i (M_{v-\varepsilon_0})+ \sum_{n=1}^{{\cal
N}_{cp}^{\nu(v)}}B(N,i(n),v-v_c^{\nu(v)},\varepsilon_0 )$.
\smallskip
\label{MainThm2}
\end{theorem}
%%%%%%%%%%%%%%%%%%%%%%%%%%%%%%%%%%%%%%%%%%%%%%%%%%%%%%%%%%%%%%%%%%%%%%%%%%%%%%

The proof of formula (\ref{meaning}) is worked out constructively.
This formula relates thermodynamic
entropy, defined in the microcanonical configurational ensemble, with
quantities of topological meaning (the Morse indexes) of the configuration
space submanifolds  $M_v=V_N^{-1}((-\infty,v])=
\{q=(q_1,\dots,q_N)\in{\Bbb R}^{N}\vert V_N(q)\leq v\}$.

After Morse theory, topology changes of the manifolds $M_v$ can be put
in one-to-one correspondence with the existence of critical points of the
potential function $V_N(q_1,\dots,q_N)$. A point $q_c$ is a
{\it critical point} if $\nabla V_N(q)\vert_{q=q_c}=0$. The potential
energy value $v_c=V_N(q_c)$ is said to be a {\it critical value} for the
potential function. Passing a critical value $v_c$, the manifolds
$M_v$ change topology. Within the framework of Morse theory, if the potential
$V_N$ is a {\it good Morse function}, that is a regular function
bounded below and with non-degenerate Hessian (that is the Hessian has no
vanishing eigenvalue), then topology changes occur through the
{\it attachment of handles} in the neighborhoods of the critical
points. Therefore, in order to establish the relationship between
entropy and configuration space topology, we have to unfold the contribution
given to the volume of $M_v$ by suitably defined neighborhoods of all the
critical points contained in $M_v$ because it is within these neighborhoods
that the relevant information about topology is contained.

This result is made possible by the idea of exploiting the existence of the
so-called Morse chart in the neighborhood of any nondegenerate critical point
of the potential function $V_N$. In fact, the Morse chart allows to represent
the
{\it local} analytic form of the equipotential hypersurfaces in an universal
form independent of the potential energy value at the critical point, and
only dependent upon the index of the critical point (equal to the number of
negative eigenvalues of the Hessian of the potential) and, obviously, upon the
dimension $N$ of configuration space.
Hence the possibility of a formal computation of the contribution of the
neighborhoods of all the critical points to the volume of $M_v$ as a function
of $v$.

\begin{pro}
Let us consider the definition of the configurational
microcanonical entropy  $S^{(-)}_N(v)$, already given in Eq.(\ref{pallaM}),
\begin{equation}
S^{(-)}_N(v)=\frac{1}{N} \log M(v, N)~,
\label{app-entr}
\end{equation}
with
\begin{eqnarray}
M (v,N) =\int_{V_N(q)\leq v}\ d^Nq &= &\int_0^v d\eta
\int_{(\Lambda^d)^{\times n}} d^Nq\ \delta [V_N(q) - \eta] \nonumber\\
&=&\int_0^v d\eta \  \int_{\Sigma_\eta}\ \frac{d\sigma}{\Vert \nabla
V_N\Vert}~, \label{volconf}
\end{eqnarray}
where we have set equal to zero the minimum of the potential.
Let $\{ \Sigma_{v_c^j}\}$ be the family of all the critical level
sets (in general not manifolds) of the potential, that is the
constant potential energy hypersurfaces that contain at least one
{\it critical point} $q^{(i)}_c$, where
$\nabla V_N(q)\vert_{q=q^{(i)}_c}=0$.
For a potential which is a good Morse function, after the Sard
Theorem (see Corollary 2 at p.200 of Ref.\cite{palais}), at any finite
dimension $N$, and below any
finite upper bound of the potential energy, the number of critical
points in configuration space and thus also the set of critical
values $\{ v_c^j\}_{j\in{\Bbb N}}$, are {\it finite, isolated} and
such that $v_c^j< v_c^k$ if $j<k$, so
that any energy interval  $v_0 \leq v \leq v_1$ is the union of a
finite number of open intervals free of critical points.

In order to split the integration on $M_v$ into two parts: the integration on
the union of the neighborhoods of all the critical points contained in $M_v$
and the integration on its complement in $M_v$, we have defined for each
critical point $q^{(i)}_c$ its {\it pseudo-cylindrical} neighborhood
$\Gamma (q^{(i)}_c,\varepsilon_0)$;
$\varepsilon_0$ is the thickness -- in potential energy -- of the neighborhood.
The assumption that the number of critical values $\vb_c^j$ is at most
linearly growing with $N$ entails, together with Sard theorem, that
$\varepsilon_0$ is finite.

%\medskip
\noindent Let us now split the integration on $M_v$ into  the integration on
$M_v \cap \bigcup_i\Gamma (q^{(i)}_c,\varepsilon_0)$ and on its complement
$M_v \setminus\bigcup_i\Gamma (q^{(i)}_c,\varepsilon_0)$.
We have
\begin{equation}
\int_{M_v} \ d^Nq =
\int_{M_v \setminus\bigcup_{i=1}^{{\cal N}(v+\varepsilon_0)}\Gamma (q^{(i)}_c,\varepsilon_0)}
\ d^Nq ~ + ~
\int_{M_v \cap\bigcup_{i=1}^{{\cal N}(v+\varepsilon_0)}\Gamma (q^{(i)}_c,\varepsilon_0)}
\ d^Nq~~,
\label{splitvol}
\end{equation}
where ${\cal N}(v)$ is the number of critical points of $V_N(q)$ up to the
level $v$. We can equivalently write
\begin{equation}
{\rm vol}(M_v)=\int_{M_v \setminus\bigcup_{i=1}^{{\cal N}(v+\varepsilon_0)}
\Gamma (q^{(i)}_c,\varepsilon_0) } \ d^Nq +
\sum_{j=1}^{{\cal N}_{cl}(v+\varepsilon_0)}\sum_{m=1}^{{\cal N}_{cp}^j}
\int_{M_v \cap \Gamma_j(q^{(m)}_c,\varepsilon_0)}\ d^Nq
\label{split}
\end{equation}
where ${\cal N}_{cl}(v)$ is the number of critical levels $\Sigma_{v_c^j}$
such that
$v_c^j< v$, and ${\cal N}_{cp}^j$ is the number of critical points on the
critical hypersurface $\Sigma_{v_c}^j$ and where we have changed the notation
of the pseudo-cylindrical neighborhoods to $\Gamma_j(q^{(m)}_c,\varepsilon_0)$
labelling with $j$ the level set to which it belongs and numbering with $m$
the critical points on the $j-$th level set. Notice that
${\cal N}(v)=\sum_{j=1}^{{\cal N}_{cl}(v)}{\cal N}_{cp}^j$.
%v_c^{\nu(v)+1}

Then we use the co-area formula in the r.h.s. of Eq.(\ref{volconf}) to rewrite
Eq.(\ref{split}); a distinction is necessary between two cases for
$\Sigma_v=\partial M_v$: its label $v$ is closer than $\varepsilon_0$ to a
critical level or not; thus we obtain
\begin{eqnarray}
{\rm vol}(M_v)&=&\int_{M_v \setminus\bigcup_{i=1}^{{\cal N}(v+\varepsilon_0)}
\Gamma (q^{(i)}_c,\varepsilon_0)} \ d^Nq \nonumber\\
&+& \sum_{j=1}^{{\cal
N}_{cl}(v+\varepsilon_0)}\sum_{m=1}^{{\cal N}_{cp}^j}
\int_{v_c^j-\varepsilon_0}^{v_c^j+\varepsilon_0}\ d\eta \
\int_{\Gamma_j(q^{(m)}_c,\varepsilon_0)} d^Nq\ \delta [V_N(q) -
\eta ]
%\int_{\Sigma_\eta\cap \Gamma_j(q^{(m)}_c,\varepsilon_0)}
% \frac{d\sigma} {\Vert\nabla v_c^{\nu(v)+1}V_N\Vert}
\label{split1}
\end{eqnarray}
when $v> v^{\nu(v)}_c+\varepsilon_0$ and $v< v_c^{\nu(v)+1}-\varepsilon_0$, where
%$\nu(v)=\max \{ j\vert v_c^j\leq v\}$; whereas
$\nu(v)$ is such that $v^{\nu(v)}_c < v < v_c^{\nu(v)+1}$; whereas
\begin{eqnarray}
{\rm vol}(M_v)&=&\int_{M_v \setminus\bigcup_{i=1}^{{\cal N}(v+\varepsilon_0)}
\Gamma (q^{(i)}_c,\varepsilon_0)} \ d^Nq\nonumber\\
& +& \sum_{j=1}^{{\cal
N}_{cl}(v)-1}\sum_{m=1}^{{\cal N}_{cp}^j}
\int_{v_c^j-\varepsilon_0}^{v_c^j+\varepsilon_0}\ d\eta
\int_{\Gamma_j(q^{(m)}_c,\varepsilon_0)} d^Nq\ \delta [V_N(q) -
\eta ]
%\int_{\Sigma_\eta\cap \Gamma_j(q^{(m)}_c,\varepsilon_0)}
% \frac{d\sigma} {\Vert\nabla v_c^{\nu(v)+1}V_N\Vert}
\nonumber\\
&+&\sum_{m=1}^{{\cal N}_{cp}^{\nu(v+\varepsilon_0)}}
\int_{v_c^{\nu(v)}-\varepsilon_0}^{v}\ d\eta
\int_{\Gamma_{\nu(v)}(q^{(m)}_c,\varepsilon_0)} d^Nq\ \delta
[V_N(q) - \eta ]
%\int_{\Sigma_\eta\cap \Gamma_{\nu(v)}(q^{(m)}_c,\varepsilon_0)}
% \frac{d\sigma} {\Vert\nabla V_N\Vert}
~.
\label{splitt1}
\end{eqnarray}
when $ v^{\nu(v)+1}_c-\varepsilon_0< v$ or $v< v^{\nu(v)}_c+\varepsilon_0$.

Near to any critical point, a second order power series expansion
of $V(q)$ reads
\[
V_N^{(2)}(q)=
V_N(q_c)+\frac{1}{2}\sum_{i,j}\frac{\partial^2V_N}{\partial
q_i\partial q_j}\ (q^i-q_c^i)\ (q^j-q_c^j)
\]
For sufficiently small $\varepsilon_0$, the integrals
$\int_{\Gamma_j(q_c,\varepsilon_0)} d^Nq\ \delta [V_N(q) - \eta ]$
can be replaced with arbitrary precision by
$\int_{\Gamma_j(q_c,\varepsilon_0)} d^Nq\ \delta [V_N^{(2)}(q) -
\eta ]$. Moreover, if $V_N(q)$ is a good Morse function, then a
coordinate transformation  exists to the so-called {\it Morse
chart} \cite{hirsch} such that
\[
{\widetilde V}_N^{(2)}(x)= V_N(q_c)- \sum_{l=1}^{k} x_l^2 +
\sum_{l=k+1}^{N} x_l^2
\]
where $k$ is the Morse index of $q_c$. Using Morse chart we have
\begin{equation}
\int_{\Gamma_j(q_c,\varepsilon_0)} d^Nq\ \delta [V^{(2)}_N(q)
-\eta ] = \int_{\Gamma_j(q_c,\varepsilon_0)} d^Nx\ \vert \det
J\vert \ \delta [{\widetilde V}^{(2)}_N(x) - \eta ]
\label{deltinteg}
\end{equation}
where $J$ is the Jacobian of the coordinate transformation.

Using Morse coordinates  inside the pseudo-cylinder ${
\Gamma_j}(q^{(m)}_c,\varepsilon_0)$ around the critical point
$q^{(m)}_c$, we see that each part of an hypersurface $\Si_\eta
\cap{\Gamma_j}(q^{(m)}_c,\varepsilon_0)$ is a quadric
\bena
\xi =
\eta - v_c^j = - \sum_{l=1}^{k_m} x_l^2 + \sum_{l=k_m+1}^{N} x_l^2
= - \mid X \mid^2 + \mid Y \mid^2 ~~,
\label{conique}
\eena
where the Morse index of $q_c^{(m)}$ is denoted by $k_m$, so that $\mid X
\mid^2=\sum_{l=1}^{k_m} x_m^2$ and $\mid Y \mid^2 =
\sum_{l=k_m+1}^{N} x_l^2 $.
Thus we rewrite the r.h.s. of Eq.(\ref{deltinteg}) as
\bena
&&\!\!\!\vert \det J\vert \int_{\Gamma_j(q^{(m)}_c,\varepsilon_0)} d^Nx\  \
\delta ( -\vert X\vert^2 +\vert Y\vert^2 -\xi ) \nonumber\\
&&\!\!=\vert \det J\vert\nonumber\\
&&\int_{\Gamma_j(q^{(m)}_c,\varepsilon_0)}
\!\!\!\!\! d\Omega^{k_m-1}d\Omega^{N-k_m-1}d\vert X\vert d\vert Y\vert
 \vert X\vert^{k_m-1}
 \vert Y\vert^{N-k_m-1}\
\delta ( -\vert X\vert^2 +\vert Y\vert^2 -\xi )\nonumber\\
\label{deltinteg1}
\eena
where $d\Omega^r$ is the solid angle element in $r$ dimension, whose
integration yields the surface $C_r$ of the $r$-dimensional hypersphere
of unit radius.
Putting $z=\vert X\vert^2$ and integrating on the angular coordinates
we get
\begin{equation}
\frac{1}{2}\vert \det J\vert C_{N-k_m-1}C_{k_m-1}
\int_{0}^{\alpha(\xi,r)}
\!\!\!\! dz \ \int_{\sqrt{\xi}}^{\beta(\xi,r)}
\!\!\!\! d\vert Y\vert  \vert Y\vert^{N-k_m-1} z^{(k_m-2)/2}
\delta ( -z +\vert Y\vert^2 -\xi )
\label{deltinteg2}
\end{equation}
where $\alpha(\xi,r)=(\sqrt{\xi^2+4r^2}-\xi )/2$ and
$\beta(\xi,r)= \sqrt{(\sqrt{\xi^2+4r^2}+\xi )/2}$. These expressions stem
from the definition of $\Gamma_j(q_c,\varepsilon_0)$ whose  boundaries have
to be orthogonal to the potential level sets described by Eq.(\ref{conique}).
These boundaries are given by the equation $\vert X\vert \vert Y\vert =r$.
Putting $y=\vert Y\vert$, from Eq.(\ref{deltinteg2}) when $\xi>0$ we obtain
\begin{equation}
\frac{1}{2} J_{jm}\ C_{N-k_m-1}C_{k_m-1}
 \ \int_{\sqrt{\xi}}^{\beta(\xi,r)}
\!\!\!\! dy\ y^{N-k_m-1}\ (y^2 -\xi )^{(k_m-2)/2}
\label{deltinteg3}
\end{equation}
and when $\xi<0$ we obtain
\begin{equation}
\frac{1}{2} J_{jm}\ C_{N-k_m-1}C_{k_m-1}
 \ \int_{0}^{\beta(\xi,r)}
\!\!\!\! dy\ y^{N-k_m-1}\ (y^2 -\xi )^{(k_m-2)/2}
\label{deltinteg4}
\end{equation}
where $C_{N\!-\!k_m\!-\! 1}$ and $C_{k_m}$ are
surfaces of hyperspheres of unit radii, that is
$C_{n}=2\pi^{n/2}/(n/2-1)!$ (for $n$ even)
and $C_{n}=2^{(n+1)/2}\pi^{(n-1)/2}/(n-2)!!$ (for $n$ odd); $\ J_{jm}$ stands
for the numerical absolute value of the determinant of $J$
computed at the critical level $v_c^j$ and at the critical point $q_c^{(m)}$.
By defining
\begin{equation}
 F_+(\xi , k_m, N)=
 \ \int_{\sqrt{\xi}}^{\beta(\xi,r)}
\!\!\!\! dy\ y^{N-k_m-1}\ (y^2 -\xi )^{(k_m-2)/2}
\label{deltinteg5}
\end{equation}
and
\begin{equation}
 F_-(\xi , k_m, N)=
 \ \int_{0}^{\beta(\xi,r)}
\!\!\!\! dy\ y^{N-k_m-1}\ (y^2 -\xi )^{(k_m-2)/2}
\label{deltinteg6}
\end{equation}
we can now write
\begin{eqnarray}
{\rm vol}(M_v)&=&\int_{M_v \setminus\bigcup_{i=1}^{{\cal N}(v+\varepsilon_0)}
 \Gamma (q^{(i)}_c,\varepsilon_0 ) } d^Nq \nonumber\\
&+&
\sum_{j=1}^{{\cal N}_{cl}(v+\varepsilon_0)}\sum_{m=1}^{{\cal N}_{cp}^j}\frac{1}{2}
C_{N\!-\!k_m\!-\! 1}C_{k_m\! -\! 1} J_{jm}
\int_{-\varepsilon_0}^{\varepsilon_0}\
d\xi \ F(\xi , k_m, N).
%\nonumber\\
%&&
\label{split2}
\end{eqnarray}
when $v> v^{\nu(v)}_c+\varepsilon_0$ and $v< v_c^{\nu(v)+1}-\varepsilon_0$,
where
$\nu(v)=\max \{ j\vert v_c^j\leq v\}$, or
\begin{eqnarray}
\!\!\!{\rm vol}(M_v)&=&\int_{M_v \setminus\bigcup_{i=1}^{{\cal N}(v+\varepsilon_0)}
 \Gamma (q^{(i)}_c,\varepsilon_0 ) } d^Nq\nonumber\\
& +&
\sum_{j=1}^{{\cal N}_{cl}(v)-1}\sum_{m=1}^{{\cal N}_{cp}^j}\frac{1}{2}
C_{N\!-\!k_m\!-\! 1}C_{k_m\! -\! 1}  J_{jm}
\int_{-\varepsilon_0}^{\varepsilon_0}\
d\xi \ F(\xi , k_m, N)\nonumber\\
&+&\sum_{m=1}^{{\cal N}_{cp}^{\nu(v+\varepsilon_0)}}\frac{1}{2}
C_{N\!-\!k_m\!-\! 1}C_{k_m\!-\! 1} J_{{\tilde j}m}
\int_{-\varepsilon_0}^{v-v_c^{\nu(v)}}\
d\xi \ F(\xi , k_m, N)
\label{splitt2}
\end{eqnarray}
with ${\tilde j}={\cal N}_{cl}(v)$, when $ v^{\nu(v)+1}_c-\varepsilon_0< v$
or $v< v^{\nu(v)}_c+\varepsilon_0$. In Eqs.(\ref{split2}) and (\ref{splitt2})
we have put
$\int_{-\varepsilon_0}^{\varepsilon_0}\ d\xi \ F(\xi , k_m, N)=
\int_{-\varepsilon_0}^{0}\ d\xi \ F_-(\xi , k_m, N) +
\int_{0}^{\varepsilon_0}\ d\xi \ F_+(\xi , k_m, N)$.

Notice that it is
${\cal N}(v)=\sum_{i=0}^N \mu_i (M_v)$, where $\mu_i (M_v)$ are the
multiplicities
of the critical points of index $i$ (there are at most $N+1$ values for
the indexes
of critical points at dimension $N$) below the energy value $v$.

Therefore, we can rearrange the double summation in Eqs.(\ref{split2}),
(\ref{splitt2}) by
expressing it as a double summation on all the possible values of the Morse
indexes and on the number of critical points for each value of the Morse
index, that is
\begin{equation}
\sum_{i=0}^N\sum_{k=1}^{\mu_i(M_v)}A(N,i,\varepsilon_0) J_{j(i,k) m(i,k)}
\label{sommaindici}
\end{equation}
where, since the integrals in Eqs.(\ref{split2}), (\ref{splitt2}) are
independent of
the index $j$, we have defined a set of positive coefficients
$A(N,i,\varepsilon_0 )$ as
\begin{equation}
A(N,i,\varepsilon_0 )=\frac{1}{2} C_{N\!-\!i\!-\!
1}C_{i}\int_{-\varepsilon_0}^{\varepsilon_0}\ d\xi\
F(\xi , i, N)\ .
\end{equation}
We remark that the term (\ref{sommaindici}), being a function of the Morse indexes
$\mu_i(M_v)$, has topological meaning. In order to make clearer this topological meaning,
we rewrite (\ref{sommaindici}) in the equivalent form of a weighed sum of Morse indexes
as follows.
From the set of positive numbers $J_{j(i,k) m(i,k)}$ we define
\begin{equation}
g_i =\frac{1}{\mu_i(M_v)}\sum_{k=1}^{\mu_i(M_v)} J_{j(i,k) m(i,k)}
\end{equation}
and rewrite the second term of the r.h.s. of Eq.(\ref{split2}) as
\begin{equation}
\sum_{i=0}^N A(N,i,\varepsilon_0 )\ g_i\ \mu_i (M_v). \label{Amu}
\end{equation}

Equation (\ref{sommaindici}) is here recast in a form which makes
somewhat clearer its dependence on Morse indexes.
Moreover, we introduce also the coefficients
\begin{equation}
B(N,i,v-v_c^{\nu(v)},\varepsilon_0 )=\frac{1}{2} C_{N\!-\!i\!-\!
1}C_{i} J_{{\tilde j} m(i,k(i))}
\int_{-\varepsilon_0}^{v-v_c^{\nu(v)}}\ d\xi\ F(\xi , i, N)\ ,
\end{equation}
where $k(i)$ stems from $j(i,k)={\tilde j}$,
such that for $v=v_c^{\nu(v)}$ it is $B(N,i,0,\varepsilon_0 )=0$, and
for $v-v_c^{\nu(v)}=\varepsilon_0$ it is
$B(N,i,\varepsilon_0,\varepsilon_0 )=A(N,i,\varepsilon_0 )g_i$.

For the purposes of the present proof, we are not concerned about
the complication of the coefficients $A(N,i,\varepsilon_0 )$ and
$B(N,i,v-v_c^{\nu(v)},\varepsilon_0 )$ because all what we need, in order to
make the link between configuration space topology and
thermodynamics, is that the second term in the volume splitting in
Eq.(\ref{splitvol}) can be written in the form (\ref{Amu}). In
fact, now we can write the entropy per degree of freedom as
\begin{eqnarray}
S^{(-)}_N(v)&=& \frac{1}{N} \log M(v, N)= \frac{1}{N} \log \int_{M_v} d^N q
\label{entropy40}\\
&=& \frac{1}{N} \log \left[ \int_{M_v
\setminus\bigcup_{i=1}^{{\cal N}(v+\varepsilon_0)}
\Gamma(q^{(i)}_c,\varepsilon_0)}\ d^Nq + \int_{M_v
\cap\bigcup_{i=1}^{{\cal N}(v+\varepsilon_0)}\Gamma(q^{(i)}_c,\varepsilon_0)}
\ d^Nq
\right] \nonumber\\
&=&\frac{1}{N} \log \left[ \int_{M_v \setminus\bigcup_{i=1}^{{\cal
N}(v)} \Gamma(q^{(i)}_c,\varepsilon_0)}\ d^Nq + \sum_{i=0}^N
A(N,i,\varepsilon_0 ) g_i \ \mu_i (M_v)
  \right] \nonumber \ ,
%\label{entropy40}
\end{eqnarray}
when $v<v_c^{\nu(v) +1}-\varepsilon_0$ or $v>v_c^{\nu(v) }+\varepsilon_0$, or
\begin{eqnarray}
S^{(-)}_N(v) &=&\frac{1}{N} \log \left[ \int_{M_v
\setminus\bigcup_{i=1}^{{\cal N}(v+\varepsilon_0)}
\Gamma(q^{(i)}_c,\varepsilon_0)}\ d^Nq + \sum_{i=0}^N
A(N,i,\varepsilon_0 ) \ g_i \mu_i (M_{v-\varepsilon_0})\right. \nonumber\\
&+&\left. \sum_{n=1}^{{\cal
N}_{cp}^{\nu(v+\varepsilon_0)+1}}B(N,i(n),v-v_c^{\nu(v)},\varepsilon_0 )
  \right] \ ,
\label{entropyx}
\end{eqnarray}
when $v>v_c^{\nu(v) +1}-\varepsilon_0$ or $v<v_c^{\nu(v) }+\varepsilon_0$.

The equation above links thermodynamic entropy with the Morse
indexes of the configuration space submanifolds $M_v$, that is
with their topology. In fact, according to Bott's ``critical-neck
theorem'' \cite{bott}, any change with $v$ of any index $\mu_i
(M_v)$, $i=0,\dots,N$, which can only be due to the crossing of a
critical level, is associated with a topology change of the $M_v$.

Conversely, any topology change, in the sense of a loss of
diffeomorphicity, occurring to the $M_v$ when $v$ is varied, is
signaled by one or more changes of the Morse indexes $\mu_i
(M_v)$ because, after the ``non-critical neck theorem''
\cite{palais}, this has to be the consequence of the crossing of a
critical level.

We remark that the two terms $\sum A_i g_i \mu_i$ and $\sum_i B_i$
in Eq.(\ref{entropyx}) stem from the same term in the volume splitting
(\ref{splitvol}) (the union of neighborhoods of critical points), so that
they will both participate in producing the development of singularities
proper to a phase transition (for more details about the role of these
two terms, see Remark \ref{remark2}).

Let us now show that the coefficients $B(N,i,v-v_c^{\nu(v)},\varepsilon_0 )$
are smooth functions of $v$ at any finite $N$.

Noting that $d^k B(N,i,v,\varepsilon_0)/dv^k =
d^{(k-1)} F(\xi,k_m,N)/d\xi^{(k-1)}$, we focus on the smoothness of
$F_\pm (\xi,k_m,N)$. Let us consider $F_+(\xi,k_m,N)$. There are two cases:
$k_m$ is even so that $(k_m-2)/2$ is an integer; $k_m$ is odd so that
$(k_m-2)/2 = n + 1/2$ with $n\in\{-1\}\cup{\Bbb N}$. In the first case,
by iteratively applying the derivation formula
$
\frac{d}{d\alpha} \int_{\phi(\alpha)}^{\psi(\alpha)}dx f(x,\alpha)=
\frac{d\psi(\alpha)}{d\alpha}f(\psi(\alpha),\alpha)- \frac{d\phi(\alpha)}
{d\alpha}
f(\phi(\alpha),\alpha)+  \int_{\phi(\alpha)}^{\psi(\alpha)}dx \frac{\partial
f(x,\alpha)}{\partial\alpha}
$ to  Eq. (\ref{deltinteg5}), one immediately realizes that $F(\xi,k_m,N)$
is smooth.

In the second case, Eq.(\ref{deltinteg5}) is rewritten as
\begin{equation}
 {\cal I}(\xi ,n, N) =
 \ \int_{\sqrt{\xi}}^{\beta(\xi,r)}
\!\!\!\! dy\ y^{N-k_m-1}\ (y^2 -\xi )^{n+\frac{1}{2}}
\label{calI}
\end{equation}
which, after integration by parts, yields the recursion formula
\begin{equation}
{\cal I}(\xi ,n, N) =\left.
\frac{(y^2-\xi)^{n+\frac{3}{2}}y^{N-1}}{2+2n+N}
\right\vert_{y=\beta(\xi,r)} + \frac{(N-1)\xi}{2+2n+N}\
{\cal I}(\xi ,n, N-2)\ .
\label{recursion}
\end{equation}
For this recursion formula there are two possible initial conditions. The
first initial condition is obtained by direct integration of
${\cal I}(\xi ,n, N)$ for $N=k_m+2$ and $n>0$, and is
\begin{equation}
{\cal I}(\xi ,n, N=k_m+2) = \left. \frac{(y^2-\xi)^{n+\frac{3}{2}}}
{3+2n}\right\vert_{y=\beta(\xi,r)}\ ;
\label{acif}
\end{equation}
the second initial condition for the recursion (\ref{recursion})
is obtained by working out ${\cal I}(\xi ,n, N)$ for $N=k_m+1$ and $n>0$.
From the above definition of ${\cal I}(\xi ,n, N)$ we get
\begin{eqnarray} {\cal I}(\xi ,n, N=k_m+1) &= &\left.
\frac{(y^2-\xi)^{n+\frac{3}{2}}} {2(n+1)
y}\right\vert_{y=\beta(\xi,r)} - \frac{\xi}{2(n+1)}\
\int_{\sqrt{\xi}}^{\beta(\xi,r)} dy
\frac{(y^2-\xi)^{n+\frac{1}{2}}}{y^2}\nonumber\\ & &
\label{integrale1}
\end{eqnarray}
 where the integral in the r.h.s. -- that
we denote by $I(n)$ -- can be solved by introducing the  double
recursion
%\be
\begin{gather}
I(n)=\left.\frac{2(y^2-\xi)^{n+\frac{3}{2}}}{(4n+3)y^3}
\right\vert_{y=\beta(\xi,r)}
- \frac{3\xi}{4n+3} H(n)
\nonumber\\
H(n)= I(n-1) - \xi H(n-1)
\label{doublerecur}
\end{gather}
%\ee
which leads to
\be H(n+1)=
\left.\frac{2(y^2-\xi)^{n+\frac{3}{2}}}{(4n+3)y^3}
\right\vert_{y=\beta(\xi,r)} - \left( \frac{3\xi}{4n+3} + \xi\right)
H(n) \label{accaenne} \ee
 together with \be H(1) = \left.\left( \frac{\xi}{3y^3} -
\frac{4}{3y}\right) \sqrt{y^2-\xi} + \log (y+ \sqrt{y^2 -
\xi})\right\vert_{y=\beta(\xi,r)}- \log (\sqrt{\xi}) .
\label{acca1}
\ee

From Eq. (\ref{acca1}) and by Eq. (\ref{accaenne}) the system of
double recursion (\ref{doublerecur}) is solved. From this we get
the required second initial condition to be inserted in Eq. (\ref{recursion}).
Now, notice that all the terms that are either powers of $(y^2-\xi)$ or
simply powers of
$y$, are to be computed
at $y=\beta(\xi,r)= \sqrt{(\sqrt{\xi^2+4r^2}+\xi )/2}$, so that they are
never vanishing
functions of $\xi$, as a consequence these terms are infinitely many times
differentiable
with respect to $\xi$. The apparent singularity of $H(1)$ for $\xi =0$ is
cured in
Eq. (\ref{doublerecur}) where it enters multiplied by $\xi$. We remark that
the recursion
formula given in Eq. (\ref{recursion}) holds also for $n=-1, 0$ which
correspond to
Morse indexes $k_m=1, 2$ respectively; the initial conditions for the
recursion are to
be explicitly computed from Eq. (\ref{calI}) by substituting $n=-1$ or $n=0$.
Also for these
special values of $n$, one is left with manifestly smooth functions,
provided that $N>2$,
an obviously acceptable "restriction" in our context. This concludes the proof
of the smoothness of the coefficients $B(N,i,v-v_c^{\nu(v)},\varepsilon_0 )$.
%~\qed
\medskip

%%%%%%%%%%%%%%%%%%%%%%%%%%%%%%%%%%%%%%%%%%%%%%%%%%%%%%%%%%%%%%%%%%%%%%%%%%%%%%%
Let us now come to the proof of the statement of Theorem \ref{MainThm2} which
says that  the source of a phase transition can only be the second term in
square brackets in Eq.(\ref{meaning}), which is of topological meaning.
To this purpose
we have to resort to the Main  Theorem of paper I and to its Corollary 1.

Let us assume that only one critical value $\vb_c$ exists in a given
interval $[ \vb_0,\vb_1]$.
After Sard theorem \cite{palais}, at any
finite $N$ there is a finite number of isolated critical points on
$\Sigma_{N\vb_c}$. For any arbitrarily small $\delta >0$,
the Main Theorem of paper I and its Corollary 1 still apply to
the two subintervals $[\vb_0, \vb_c-\delta]$ and $[ \vb_c+\delta
,\vb_1]$. In order to understand why a breakdown of uniform
boundedness in $N$ of $\vert\partial^kS_N^{(-)}/\partial v^k\vert$
for $k=3$ or $k=4$ can be originated only by the topological term in
r.h.s. of Eq.(\ref{meaning}), we consider each critical point
$q^{(i)}_c$ on $\Sigma_{N{\vb_c}}$ enclosed in a small
pseudo-cylindrical neighborhood $\Gamma(q^{(i)}_c,\varepsilon_0)$
of thickness ${\varepsilon}_0$ and we take
${\varepsilon}<{\varepsilon}_0$ arbitrarily close to
${\varepsilon}_0$. From Morse theory, we know that passing a
critical value $v_c$ entails that for each critical point of
index $k_i$  a $k_i$-handle $H^{N,k_i}$ is attached to $M_{v <
v_c}$ so that the following diffeomorphism holds
\begin{equation}
M_{(v_c+{\varepsilon}_0)}\approx
M_{(v_c-{\varepsilon}_0)}\bigcup_{\phi_1} H^{N,k_1}
\bigcup_{\phi_2}H^{N,k_2}\dots \bigcup_{\phi_n}H^{N,k_n}
\end{equation}
where a $k_i$-handle in $N$ dimensions ($0\leq k_i\leq N$) is the
product of two disks, a $k_i$-dimensional disk $D^{k_i}$, and
another $(N-k_i)$-dimensional disk $D^{(N-k_i)}$ s.t.
$H^{N,k_i}=D^{k_i}\times D^{(N-k_i)}$, and where $\bigcup_{\phi_i}$
stands for the attachment of $H^{N,k_i}$ to $M_{(v_c-{\varepsilon}_0)}$
through the embedding $\phi_i:{\Bbb S}^{k_i-1}\times D^{N-k_i}\rightarrow
\partial M_{(v_c-{\varepsilon}_0)}$
(where ${\Bbb S}$ is an hypersphere; details can be found in
\cite{palais,hirsch,milnor}).

The excision of the pseudo-cylindrical neighborhoods
$\Gamma(q^i_c,\varepsilon_0)$ of all the critical points
$q^{(i)}_c\in\Sigma_{N{\vb_c}}$ implies that all the manifolds
${\overline M}_{v}:=M_{v}\setminus\bigcup_{i=1}^{\#\ crit.pts.}
\Gamma(q^{(i)}_c,\varepsilon_0)$ with
$N\vb_c-{\varepsilon}_0< v < N\vb_c+{\varepsilon}_0$
are free of critical points and, consequently, are diffeomorphic. In fact,
for any $v,v^\prime\in{\Bbb R}$ such that
$N\vb_c-{\varepsilon}_0< v < v^\prime < N\vb_c+{\varepsilon}_0$,
${\overline M}_{v}$ is a deformation retraction of
${\overline M}_{v^\prime}$ through the
flow associated with the vector field $X=-\nabla V_N/\Vert\nabla V_N\Vert^2$
\cite{palais,hirsch}.

Now, defining ${\overline M}(v,N)= {\rm vol} ({\overline M}_{v})$ and
$\Gamma(v,N)= {\rm vol}[\bigcup_{i=1}^{\#\ crit.pts.}
\Gamma(q^{(i)}_c,\varepsilon_0)]$, equation (\ref{meaning}) becomes
\begin{eqnarray}
S^{(-)}_N(v)&=&\frac{1}{N} \log \left[ {\overline M}(v,N) + \Gamma(v,N)\right]
\nonumber\\
&=&\frac{1}{N} \log \left[{\overline M}(v,N)\right] +
\frac{1}{N} \log \left[ 1 + \frac{\Gamma(v,N)}{{\overline M}(v,N)}\right]~.
\label{ozzac}
%\nonumber
\end{eqnarray}
By applying the Main Theorem of paper I and its Corollary 1 to the first
term in the r.h.s. of the equation above, we know that
$\frac{1}{N}\vert\partial^k \log \left[{\overline M}(v,N)\right]
/\partial v^k\vert$ for $k=1,\dots,4$, are uniformly bounded in $N$, and thus
no phase transition can be attributed to this term.

Then, let us consider the second term of the r.h.s. of the equation above.
By computing its first derivative we obtain
\begin{equation}
\frac{d}{dv}\frac{1}{N} \log \left[ 1 + \frac{\Gamma(v,N)}{{\overline M}(v,N)}
\right] =\frac{1}{N}\frac{\Gamma^\prime}{{\overline M}+\Gamma}-
\frac{1}{N}\frac{\Gamma}{{\overline M}+\Gamma}\left(\frac{{\overline M}^\prime}
{{\overline M}}\right)~,
\label{disopra}
\end{equation}
where $({\overline M}^\prime/{\overline M})$ stands for
$[d{\overline M}(v,N)/dv]/{\overline M}(v,N)$.
After Lemma 1,
$({\overline M}^\prime/{\overline M})$ is uniformly bounded in $N$ and
therefore so does the second
term in the r.h.s. of Eq.(\ref{disopra}). Whence, if
$\vert\partial S_N^{(-)}/\partial v\vert$ were to grow with $N$ this could
not be due to the term ${\overline M}(v,N)$.

Then we compute the second derivative
\begin{eqnarray}
\frac{d^2}{dv^2}\frac{1}{N} \log \left[ 1 + \frac{\Gamma(v,N)}
{{\overline M}(v,N)}\right] &=&
\frac{1}{N}\frac{\Gamma^{\prime\prime}}{{\overline M}+\Gamma}+
\frac{1}{N}\frac{\Gamma{\overline M}({\overline M}^\prime/
{\overline M})+\Gamma\Gamma^\prime}{({\overline M}+\Gamma)^2}
\left(\frac{{\overline M}^\prime}{{\overline M}}-\frac{\Gamma^\prime}{\Gamma}
\right) \nonumber\\
&-&\frac{1}{N}\frac{\Gamma}{{\overline M}+\Gamma}
\left(\frac{{\overline M}^\prime}{{\overline M}}\right) +\frac{1}{N}\frac{d}{dv}
\left(\frac{{\overline M}^\prime}{{\overline M}}\right) ~.
\label{disopra1}
\end{eqnarray}
Again, we can observe that the uniform boundedness with $N$ of both
$({\overline M}^\prime/{\overline M})$ and $(d/dv)({\overline M}^\prime/
{\overline M})=({\overline M}^{\prime\prime}/{\overline M})-
({\overline M}^{\prime}/{\overline M})^2$ -- after Lemma 1  -- entails that if
$\vert\partial^2 S_N^{(-)}/\partial v^2
\vert$ were to grow with $N$, this could not be due to the term
${\overline M}(v,N)$.

% which are now
%diffeomorphic and free of critical points, and which still allow
%the application of Federer's derivation formula
%(\ref{federerderiv}) (see the Remark \ref{federemark} at the end of
%the Appendix),
%we know that $\vert\partial^kS_N^{(-)}/\partial v^k\vert$ for
%$k=1,\dots,4$, are uniformly bounded in $N$.

Similarly, after a lengthy but trivial computation of the third and fourth
derivatives of the second term in the r.h.s. of Eq.(\ref{ozzac}),
one finds that
${\overline M}(v,N)$ enters the various terms obtained through the ratio
${\overline M^\prime}/{\overline M}$ and through its derivatives
$\frac{d^k}{dv^k}[{\overline M}^\prime/{\overline M}]$ with
$k=1,2,3$, thus, as a consequence of Lemma 1, the uniform boundedness in $N$
of $[d{\overline M}(v,N)/dv]/{\overline M}(v,N)$ and
$\frac{d^k}{dv^k}\{[d{\overline M}(v,N)/dv]/{\overline M}(v,N)\}$ with
$k=1,2,3$, implies that if
$\vert\partial^3 S_N^{(-)}/\partial v^3\vert$ or
$\vert\partial^4 S_N^{(-)}/\partial v^4\vert$ were to grow with $N$,
this could not be due to the term
${\overline M}(v,N)$.

In conclusion, the
first term within square brackets in Eq.(\ref{meaning}) cannot be
at the origin of a phase transition, nor can it be the third one, which is
the sum of smooth functions. Only the second term of the
r.h.s. of Eq.(\ref{meaning}), which is in one-to-one correspondence
with topology changes of the $M_v$, can originate an unbound growth with $N$
of a derivative $\vert\partial^kS_N^{(-)}/\partial v^k\vert$ for some $k$, thus
entailing a phase transition.
~\qed
\end{pro}

\begin{remark}
\label{remark2}
A comment about the above considered volume splitting is in order.
At any finite $N$, the volume $M(v,N)$ is a smooth function of $v$,
as is the entropy $S^{(-)}_N(v)$.
Since the term $\sum_{i=0}^N A(N,i,\varepsilon_0 )g_i \mu_i(M_v)$
entering Eqs.(\ref{entropy40}) and (\ref{entropyx}) depends on the
integer valued functions $\mu_i(v)$, one could erroneously think that
the presence of the functions $\mu_i(v)$ in this term conflicts with
the smoothness
of volume and entropy. Of course, at finite $N$, smoothness of volume
and entropy is not lost.
In fact, the term $\sum_{i=0}^N A(N,i,\varepsilon_0 )g_i \mu_i(M_v)$
is constant in any open interval $(v_c^{\nu (v)}+\varepsilon_0,
v_c^{\nu (v)+1}-\varepsilon_0)$ and the functions
$v\rightarrowtail B(N,i,v)$ smoothly
connect in the interval $(v_c^{\nu (v)}-\varepsilon_0,
v_c^{\nu (v)}+\varepsilon_0)$ the values that the function
$\sum_{i=0}^N A(N,i,\varepsilon_0 )g_i \mu_i(M_v)$ takes in the
intervals $(v_c^{\nu (v)-1}+\varepsilon_0,
v_c^{\nu (v)}-\varepsilon_0)$ and $(v_c^{\nu (v)}+\varepsilon_0,
v_c^{\nu (v)+1}-\varepsilon_0)$.

Loosely speaking,
$\sum_{i} A(N,i,\varepsilon_0 )g_i \mu_i(M_v)+\sum_n B(N,i(n),v)$
is shaped as a "staircase" with "rounded corners".
\end{remark}

\begin{remark}
About the applicability domain of the Main Theorem proved in the present
paper, note that $V$ is required to be a finite range
potential and a good Morse function. The former is a physical assumption,
the latter a mathematical property of $V$. Finite range potentials are
typical in condensed matter systems, where even Coulomb interactions are
effective only at a finite distance because of the Debye shielding.
The mathematical property of being a good Morse function is absolutely
generic, in fact it requires that a potential function is bounded from below
and that the Hessian of the potential is nondegenerate (i.e. its eigenvalues
never vanish). Moreover, given a real-valued function $f$ of class ${\cal C}^2$
defined on an arbitrary open subset $X$ of ${\Bbb R}^N$, the mapping
$x\rightarrowtail f(x) - (a_1x_1+\dots +a_Nx_N): X\rightarrowtail {\Bbb R}$
is nondegenerate for almost all $(a_1,\dots,a_N)\in {\Bbb R}^N$ (see Chapter 6
of Ref.\cite{MorseCairns}). This means that nondegeneracy is generic whereas
degeneracy is exceptional. Continuous symmetries are the only physically relevant
source of degeneracy, however this kind of degeneracy can be removed by
adding a generic term $(a_1x_1+\dots +a_Nx_N)$ to the potential with an arbitrarily
small vector $(a_1,\dots,a_N)$. This removal of degeneracy is a rephrasing,
within the framework of Morse theory, of a standard procedure undertaken in statistical
mechanics to explicitly break a continuous symmetry, that is the addition of
an external field
whose limit to zero is taken after the limit $N\to\infty$.

\end{remark}

\begin{remark}
Let us briefly compare meaning and strength  of the Main Theorems of papers
I and II, which we denote by MT-I and MT-II respectively.
The proof of MT-I is preliminary to, and independent of, the proof of MT-II.
On the other hand, it is important to note that the hypotheses under which
MT-II applies
are fulfilled by a very broad class of physically meaningful potentials,
at variance with the case of MT-I. In fact, MT-I applies in presence of a
potential energy density interval of finite length which is definitevely (that is
asymptotically in $N$) free of critical points, a rather restrictive assumption which we
could hardly relax in the present demonstration scheme. On the contrary, we can approach
the problem from a complementary point of view allowing the existence of
critical points of the potential -- as is assumed in the hypotheses of MT-II -- with the only
limitation to a linear growth with $N$ of the number of critical values of
$V$ in a potential energy density interval of finite length. This
assumption is suggested by what happens in lattice systems where critical
levels are separated by a finite minimum energy amount, as is the case
of "spin flips" in the $1d$-XY model \cite{xymf}, mean-field XY model \cite{xymf},
p-trig model \cite{ptrig}, or of elementary configurational changes that, in
lattices and fluids, correspond to the appearance of a new critical value of $V$
at an energetic cost which is independent of $N$.

It would be impossible to prove MT-II without resorting to MT-I, though
MT-II is much stronger than MT-I. Actually, the link established by MT-II in
Eq.(\ref{meaning}), between thermodynamic entropy and topology,
tells that only through suitable variations with $N$ of
the $v$-patterns of $\sum_{i=0}^N A(N,i,\varepsilon_0 )g_i \mu_i
(M_v)$ some of the derivatives $\vert\partial^kS_N^{(-)}/\partial
v^k\vert$ can cease to be uniformly bounded in $N$ from above, thus
giving rise to a phase transition.
This is an important hint for a future rigorous investigation of
the sufficiency conditions for both MT-I and MT-II.
The combination of MT-I and MT-II
provides clear evidence of  the relevance of topology for
the phenomenon of phase transitions.

\end{remark}

%Though this problem of {\it sufficiency} is
%still wide open, we already have some useful hints provided by the exact
%analytic computation of the Euler characteristic of the submanifolds
%$M_v=\{q_1,\dots,q_N\in\Lambda^{\times N}\vert V(q_1,\dots,q_N)\leq v\}$
%for two models
%undergoing first or second order phase transitions or no phase transitions
%at all\cite{xymf,ptrig}. These results, together with the numerically computed
%Euler characteristic $\chi (\Sigma_v)~{\rm vs.}~v$ for a two-dimensional
%lattice
%$\varphi^4$ model undergoing a symmetry-breaking phase transition \cite{top2},
%suggest that phase transitions would correspond to abrupt transitions in the
%way topology changes as a function of $v$. In the so-called mean-field XY
%model, for example, the phase transition stems from the simultaneous attachment
%of handles of ${\cal O}(N)$ different types on the same critical level
%\cite{xymf}.
%%%%%%%%%%%%%%%%%%%%%%%%%%%%%%%%%%%%%%%%%%%%%%%%%%%%%%%%%%%%%%%%%%%%%%%%%%%%%%

%%%%%%%%%%%%%%%%%%%%%%%%%%%%%%%%%%%%%%%%%%%%%%%%%%%%%%%%%%%%%%%%%%%%%%%%%%%%%%
\section{Final remarks}
Let us conclude with a few general comments.
Earlier attempts at introducing topological concepts in statistical
mechanics concentrated on {\it macroscopic} low-dimensional parameter spaces.
Actually this happened after Thom's remark that the critical point shown by
the van der Waals equation corresponds to the Riemann-Hugoniot
catastrophe \cite{thom}. Hence some applications of the theory of
singularities of differentiable maps to the study of phase transitions
followed \cite{Poston}.
An elegant formulation of phase transitions as due to a topological
change of some abstract manifold of macroscopic variables was obtained by
using the Atiyah-Singer index theorem \cite{rasetti,rasetti1} and deserves
special attention because it applies to the 2$d$ Ising model, whose phase
transition is associated with a jump of the Atiyah index of some suitable
vector bundle.
This shows that also for discrete variables systems, like spin systems,
topological concepts can be useful in the study of phase transitions, provided
that the relevant manifolds are identified.

The Main Theorem, that we have proved above, makes a new kind of link between
the study of phase transitions and differential topology.
In fact, in the present work we deal with the high-dimensional
{\it microscopic} configuration space of a physical system. The level sets of
the microscopic interaction potential among  the particles -- or the regions
of configuration space bounded by them -- are the
configuration space submanifolds that necessarily have to change their
topology in correspondence with a phase transition point. The topology changes
implied here are those described within the framework of Morse theory through
{\it attachment of handles} \cite{hirsch}.

We have explicitly investigated these topology changes in some particular models.
The results so far obtained, already reported in the literature, are: {\it
i)} numerical results on the lattice $\varphi^4$ model
\cite{top2}, {\it ii)} exact analytical results on the two models
considered in Refs.\cite{xymf,ptrig}, {\it iii)}  analytical results on
the coupled rotators model. Although the models in item {\it (ii)}
do not fulfil the condition of short-range interactions assumed by MT-I and
MT-II in their present formulations, we already get a coherent scenario
illustrating in practice how topology changes in presence and in absence of
phase transitions. In fact, we observe that when the $\vb$-pattern
of the Euler characteristic $\chi(\vb )$ -- which is the probe that we use
to detect the variations of topology with $\vb$ -- approaches a smooth curve as
$N$ increases, then phase transitions are absent; this is the case
of $\chi(\Sigma_{v})$ for the one-dimensional $\varphi^4$ model
and of $\chi(M_{v})$ for the one-dimensional chain of coupled
rotators. At variance, sharp jumps or ``cuspy'' $\vb$-patterns of
$\chi(\Sigma_{v})$ or of $\chi(M_{v})$ are associated with
phase transitions, in qualitative agreement with what is expected after
MT-II; in Ref.\cite{ptrig} it is also shown that first
and second order phase transitions are signaled by markedly
different $\vb$-patterns of $\chi(M_{v})$.

Notice that in our approach the role of the potential $V$
is twofold: it determines the relevant submanifolds of configuration space
and it is a good Morse function on the same space. However, for example, in
the case of entropy driven phase transitions occurring in hard sphere gases,
the fact that the (singular) interaction potential cannot play any longer the
role of Morse function does not mean that the connection between topology and
phase transitions is lost, it rather means that other Morse functions are to
be used.
Just to give an idea of what a good Morse function could be in this case, let
us think of the sum of all the pairwise euclidean distances between the hard
spheres of a system: it is real valued, it has a minimum when the density is
maximum, that is for close packing, meaning that this function is bounded
below. The discussion of non-degeneracy is more involved and here would be out
of place, let us simply remark that Morse functions are dense and degeneracy
is easily removed when necessary.

The topology of configuration space submanifolds makes also a subtle link
between dynamics and thermodynamics because it affects both of them, the former
because it can be seen as the geodesic flow of a suitable Riemannian metric
endowing configuration space \cite{physrep}, the latter because an analytic
(though approximate) relation between thermodynamic entropy and Morse indexes
of the critical points of configuration space submanifolds can be worked out
\cite{xymf}.

Moreover, there are ``exotic'' kinds of transitional phenomena in statistical
physics, like the glassy transition of amorphous systems to a supercooled
liquid regime, or the folding transitions in polymers and proteins,
which are qualitatively unified through the so-called {\it landscape paradigm}
\cite{elandscape,sastry}
which is based on the idea that the relevant physics of these systems can be
understood through the study of the properties of the potential energy
hypersurfaces and, in particular, of their stationary points, usually called
``saddles''.  That this landscape paradigm naturally goes toward a link with
Morse theory and topology has been hitherto overlooked.
However, though at present our Main Theorem only applies to first and second
order phase transitions, the topological approach seems to have the
potentiality of unifying the mathematical description of very different kinds
 of phase transitions.

\section{Acknowledgments}
The authors wish to thank L. Casetti and S. Schreiber for comments and suggestions.
A particularly warm acknowledgment is addressed to G. Vezzosi for his
continuous interest in our work and for many helpful discussions and
suggestions.

%=========================================================================

%=========================================================================


\begin{thebibliography}{99}

\bibitem{onsager} L. Onsager, Phys. Rev. \textbf{65}, (1944) 117.

\bibitem{YLthm} C.N. Yang and T.D. Lee, Phys. Rev.\textbf{87}, (1952) 404.

\bibitem{ruelleTD} D. Ruelle, \textit{Thermodynamic formalism}, Encyclopaedia
of Mathematics and its Applications, (Addison-Wesley, New York, 1978).

\bibitem{georgii} H.O. Georgii, \textit {Gibbs Measures and Phase Transitions},
              (Walter de Gruyter, Berlin, 1988).

\bibitem{cccp} L. Caiani, L. Casetti, C. Clementi and M. Pettini,
               Phys. Rev. Lett. \textbf{79}, (1997) 4361.

\bibitem{top1} R. Franzosi, L. Casetti, L.Spinelli and M. Pettini,
               Phys. Rev. E\textbf{60}, (1999) R5009.

\bibitem{top2} R. Franzosi, M. Pettini, and L.Spinelli,
               Phys. Rev. Lett. \textbf{84}, (2000) 2774.

\bibitem{top3} L. Casetti, E.G.D. Cohen and M. Pettini,
               Phys. Rev. Lett. \textbf{82}, (1999) 4160.

\bibitem{physrep} L. Casetti, M. Pettini, and E.G.D. Cohen,
               Phys. Rep. \textbf{337}, (2000) 237-341.

\bibitem{xymf} L. Casetti, M. Pettini, and E.G.D. Cohen,
               J. Stat. Phys. \textbf{111}, (2003) 1091.

\bibitem{ptrig} L. Angelani, L. Casetti, M. Pettini, G. Ruocco, and F. Zamponi,
               Europhys. Lett.\textbf{62}, (2003) 775; Phys. Rev. E\textbf{71},
                (2005) 036152.

\bibitem{paperI}  R. Franzosi, M. Pettini, and L.Spinelli,
\textit{Topology and Phase Transitions I. Preliminary Results},
             archived in: math-ph/0505057.

\bibitem{pirl} R. Franzosi, and M. Pettini,
               Phys. Rev. Lett. \textbf{92}, (2004) 060601.

\bibitem{MorseCairns} M. Morse and S. S. Cairns, {\it Critical Point
Theory in Global Analysis and Differential Topology},
(Academic Press, New York, 1969).

\bibitem{federer} H. Federer, \textit{Geometric Measure Theory}, (Springer,
                 New York 1969), p. 249.

\bibitem{ruelle} D. Ruelle, \textit{Statistical Mechanics.
Rigorous results}, (Benjamin, Reading, 1969).

\bibitem{palais} R.S. Palais and C. Terng, \textit{Critical Point Theory and Submanifold Geometry}, (Springer, New York 1988).

\bibitem{hirsch} M.W. Hirsch, \textit{Differential Topology},
(Springer, New York 1976).

\bibitem{milnor} J. Milnor, \textit{Morse Theory}, (Princeton University Press, Princeton, 1973).

\bibitem{laurence} P. Laurence, ZAMP \textbf{40}, (1989) 258.

\bibitem{bott} R. Bott and J. Mather, {\it Topics in Topology and
Differential Geometry}, in {\sl Battelle Rencontres},
Eds. C.M. De Witt and J.A. Wheeler, p.460.
of the Main Theorem of paper I.
The combination of the Main Theorem of paper I  and  Theorem \ref{MainThm2}
of the present paper

\bibitem{thom} R. Thom, in {\it Statistical Mechanics},
Eds. S.A. Rice,
K.F. Freed, and J.C. Light, (University of Chicago Press, 1972), p.93.

\bibitem{Poston} T. Poston and I. Stewart, {\it Catastrophe Theory
and its Applications}, (Pitman Press, London, 1978), and references therein
quoted.

\bibitem{rasetti} M. Rasetti, {\it Topological concepts in the
theory of phase transitions}, in {\it Differential Geometric
Methods in Mathematical Physics},
Ed. H.D. D\"obner (Springer-Verlag, New York, 1979).

\bibitem{rasetti1} M. Rasetti, {\it Structural Stability
in Statistical Mechanics}, in {\it Springer Tracts in
Math.}, Ed. W. G\"uttinger (Springer-Verlag, New York, 1979).

\bibitem{elandscape} F.H. Stillinger, Science \textbf{267}, (1995) 1935.

\bibitem{sastry} S. Sastry, P.G. Debenedetti and F.H. Stillinger,
                 Nature \textbf{393}, (1998) 554.



\end{thebibliography}
\end{document}